\documentclass[aps,prb,preprint,showpacs,superscriptaddress]{revtex4}
\usepackage{amsmath}
\usepackage{epsfig}

\begin{document}
%\draft
%\preprint{}

%
\title{Electronic and optical properties of ferromagnetic GaMnAs in a
 multi-band tight-binding approach}
%\title{Thermopower of a single-electron transistor with a 
%                        superconducting island}

\author{Marko Turek}
\email{marko.turek@physik.uni-regensburg.de}
\affiliation{Institut f{\"u}r Theoretische Physik, Universit{\"a}t Regensburg,
             D-93040 Regensburg, Germany}
\author{Jens Siewert}
\affiliation{Institut f{\"u}r Theoretische Physik, Universit{\"a}t Regensburg,
             D-93040 Regensburg, Germany}
%\affiliation{Dipartimento di Metodologie Fisiche e Chimiche per l'Ingegneria,
% Universita di Catania,  I-95125 Catania, Italy}
\author{Jaroslav Fabian}
\affiliation{Institut f{\"u}r Theoretische Physik, Universit{\"a}t Regensburg,
             D-93040 Regensburg, Germany}

\date{\today}

\begin{abstract}
We consider the electronic properties of ferromagnetic bulk ${\rm Ga}_{1-x}{\rm Mn}_x {\rm As}$
at zero temperature using two realistic tight-binding models, one due to Tang and Flatt{\'e}
and one due to Ma{\v s}ek. In particular, we study
the density of states, the Fermi energy, the inverse participation ratio,
and the optical conductivity with varying impurity concentration $x=0.01$~--~$0.15$.
The results are very sensitive to the assumptions
made for the on-site and hopping matrix elements of the Mn impurities.
For low concentrations, $x<0.02$, Ma{\v s}ek's model shows only small
deviations from the case of p-doped GaAs with increased number of holes
while within Tang and Flatt{\'e}'s model an impurity-band forms. 
For higher concentrations $x$, Ma{\v s}ek's model shows minor
quantitative changes in the properties we studied while
the results of the Tang and Flatt{\'e} model exhibit qualitative changes
including strong localization of eigenstates with energies close
to the band edge. These differences between the two approaches
are in particular visible in the optical conductivity, where Ma{\v s}ek's
model shows a Drude peak at zero frequency while no such peak is
observed in Tang and Flatt{\'e}'s model. Interestingly, although the
two models differ qualitatively the calculated effective optical masses of both
models are similar within the range of $0.4$~--~$1.0$ of the free electron mass.
\end{abstract}


\maketitle

%\twocolumn

%\narrowtext
%
%%%%%%%%%%%%%%%%%%%%%%%%%%%%%%%%%%%%%%%%%%%%

%%%%%%%%%%%%%%%%%%%%%%%%%%%%%%%%%%%%%%%%%%%%
\section{Introduction}
\label{sec:introduction}
%%%%%%%%%%%%%%%%%%%%%%%%%%%%%%%%%%%%%%%%%%%%
Dilute magnetic semiconductors are investigated very actively
for their potential of combining ferromagnetic with semiconducting properties.
A prominent prototype is ${\rm Ga}_{1-x}{\rm Mn}_x {\rm As}$
with Mn concentrations being typically
$x=0.01$~--~$0.15$.\cite{Ohno98,Jungwirth06a,Fabian07,Dietl08}
For these intermediate
to large concentrations critical temperatures up to $170$ K
could be observed.\cite{Jungwirth05}
The host material, GaAs, is a III-V semiconductor with a band gap of $1.5$~eV.
When the Mn impurities substitute the Ga atoms they
act as acceptors which carry a local magnetic moment caused by the
half-filled d-shell. This leads to a hole-mediated ferromagnetism.\cite{MacDonald05}
The impurity binding energy is $0.11$~eV.\cite{Linnarsson97}
If the Mn concentration exceeds a critical value $x\approx 0.01$
the impurity wavefunctions at the Fermi energy
overlap sufficiently for the material to undergo a transition towards 
a metallic state.

A widely discussed question is whether the holes reside
in an impurity band which is detached from the host valence band
or in the valence band itself, which would cause different
transport properties.\cite{Dietl08}
To this end a variety of absorption
experiments\cite{Burch06a,Katsumoto01,Nagai01,Hirakawa02} and measurements
of the band gap and chemical potential \cite{Tsuruoka02,Thomas07}
in GaMnAs were performed.
For very low concentrations $x \ll 0.01$,
an impurity band is formed with the Fermi
energy residing therein.\cite{Jungwirth07a} This picture
has also been used for the intermediate concentration range,
$x \approx 0.01$~--~$0.15$, to explain
optical measurements.\cite{Burch06a} On the other hand, there is
experimental indication that the impurity band and the valence band have completely
merged, see Ref. \onlinecite{Jungwirth07a} and references therein.
Another issue concerns the localization properties of the carriers.
It has been argued that instead of the the impurity band interpretation
the assumption of a merged valence and impurity band together with
the existence of localized states in the band tail
can explain the experiments.\cite{Jungwirth07a}

Besides the experimental efforts for a better understanding of the electronic
properties of ${\rm Ga}_{1-x}{\rm Mn}_x {\rm As}$ a wide range of theoretical
models has been developed.\cite{Jungwirth06a} Among these are first-principles
calculations \cite{Sandratskii04,Stroppa08},
effective single particle tight-binding approaches
\cite{Tang04a,Dorpe05,Sankowski06,Masek07a},
tight-binding approaches in combination
with percolation theory \cite{Bhatt02,Kaminski02,DasSarma03}, dynamical mean field
theories \cite{Majidi06},
effective theories based on $k \cdot p$ Hamiltonians
\cite{Dietl00,Dietl01,Yang03,Elsen07}, and 
large-scale Monte-Carlo studies of a real space Hamiltonian \cite{Yildrim07a}.
The microscopic tight-binding approach has the advantage that one can study
disordered systems in a non-perturbative way. The basis for this approach
are the 16 ${\rm sp}^3$ valence and conduction bands of GaAs that are approximated
very closely to experimental data throughout the entire Brillouin zone. System sizes
up to approximately 2000 atoms can be treated within
this approach with reasonable numerical effort.

In this paper we present our results concerning the electronic structure
of bulk ${\rm Ga}_{1-x}{\rm Mn}_x {\rm As}$ at zero temperature. We assume the system to be
ferromagnetic by aligning all Mn core spins into the z-direction.
Specifically, we focus on the role of substitutional Mn impurities by neglecting
interstitial disorder. The basis for our studies are well-known ${\rm sp}^3$ tight-binding
models for the host material GaAs
which include spin-orbit coupling.\cite{Chadi77a,Talwar82a} The 
disorder effects are treated in a non-perturbative
way by changing the on-site and hopping terms of those Ga sites that are
replaced by the Mn impurities. Instead of describing the effects
of the Mn impurities by the two
sp-d exchange constants $N_0\alpha$ and $N_0\beta$ only\cite{Dorpe05,Sankowski06}
we follow two more elaborate
approaches.
The first approach, by Ma{\v s}ek\cite{Masek07a,Masek07b}
(referred to as model A),
uses first principles methods to determine the tight-binding
parameters of the Mn impurities in a concentration regime 
around $10$\%. The second approach, by Tang and
Flatt{\' e}\cite{Tang04a,Kitchen06a} (model B), is based on
a fit of the physically relevant tight-binding parameters to
reproduce the binding energy of a single Mn
impurity in the host material. Both approaches result in
effective single particle tight-binding models in which
the carrier--carrier correlations are included in a mean
field way. This approximation is
justified in the intermediate to high doping regime,
i.e. $x \gtrsim 0.01$.\cite{Jungwirth07a}
Nevertheless, explicit inclusion of carrier--carrier
interactions could lead to quantitative corrections
which enhance localization effects.\cite{Dietl08a}

In section \ref{sec:tight-binding} we summarize
the two approaches of how to find the reliable parameter sets
for the inclusion of the Mn impurities
and introduce our numerical method for solving
the resulting equations in more detail.
The explicit calculation of the eigenenergies and -states allows us
to investigate the density of states and the position of the
Fermi energy. In section \ref{sec:density} we show that for
Ma{\v s}ek's model (A)
the valence band is only slightly deformed while for the model
by Tang and Flatt{\' e} (B)
an impurity band forms which starts to merge with the host valence
band at $x\approx0.01$. The spatial extension of the states around
the Fermi energy is investigated in section \ref{sec:inverse}
by means of the inverse participation ratio. We show that the
states related to the impurity band show strong localization
when their energy is close to the band edge. Based on the spectrum
and the eigenfunctions we calculate the experimentally relevant
optical conductivity. The relation of these results to the experimental
findings is discussed in section \ref{sec:conductivity}.

%%%%%%%%%%%%%%%%%%%%%%%%%%%%%%%%%%%%%%%%%%%%
\section{Tight-binding approach}
\label{sec:tight-binding}
%%%%%%%%%%%%%%%%%%%%%%%%%%%%%%%%%%%%%%%%%%%%
Our analysis within the framework of a phenomenological tight-binding
approach \cite{Slater54a,Harrison99a}
is based on an effective single particle Hamiltonian
allowing for material-specific simulations of bulk GaMnAs
systems. We considered disordered super-cells of up to 2000 atoms
with the super-cell size being limited by the available computer
resources only. In this work we focus
on substitutional disorder by changing the on-site
and hopping terms of certain Ga sites which are replaced by Mn.

In order to introduce the terminology we briefly summarize the tight-binding 
approach. Using the single particle basis,
\begin{equation}
 \label{eq:basis} \chi_{\vec k o a}(\vec r) = \frac{1}{\sqrt{S}}
 \sum\limits_{\vec R} \exp \left( {\rm i} \vec k \left[ \vec R +
 \vec t_a \right]  \right) \phi_o(\vec r - \vec t_a - \vec R) \, ,
\end{equation}
the eigenstates of the Hamiltonian can be written as
\begin{equation}
 \label{eq:wavefunction} \Psi_{\vec k}^{(b)} (\vec r)  = 
 \sum\limits_{o,a} c_{\vec k o a}^{(b)} \chi_{\vec k o a}(\vec r) .
\end{equation}
Here, $o$ labels the atomic orbital $\phi_o$, $a$ specifies the atom sitting
at the position $\vec t_a$ within the super-cell, $S$ is the total number of super-cells
and $\vec R$ points to the several super-cells that are included.
The sum over $o$ and $a$ in Eq.~\ref{eq:wavefunction}
thus runs over all orbitals of all $N$ atoms in a super-cell.
Due to the periodicity and the Bloch character of the basis states
$\chi_{\vec k o a}$ the Hamiltonian is block-diagonal with respect to the
different $\vec k$ vectors. The matrix elements in this basis thus read
\begin{equation}
 \label{eq:hamiltonian} \left[ H(\vec k) \right]_{oa,o'a'}
 = \sum\limits_{\Delta \vec R_{nn}}
 \exp \left( {\rm i} \vec k \vec r_{nn} \right)
 \left\langle \phi_{oa}(\vec r) | H | \phi_{o'a'}(\vec r - \vec r_{nn})
 \right\rangle \, ,
\end{equation}
where the sum goes over all the super-cells that contain
the nearest or next to
nearest neighbors, depending on the approximation employed.
The vector $\vec r_{nn}$ points from atom $a$ to atom
$a'$ so that $\vec r_{nn} = \Delta \vec R_{nn} + \vec t_{a'} - \vec t_a$.
On-site terms are thus characterized by $\vec r_{nn} = 0$ while
hopping terms correspond to $\vec r_{nn} \neq 0$.
Numerical diagonalization of the matrix (\ref{eq:hamiltonian}) returns the
eigenenergies and -vectors as a linear combination of the localized atomic
orbitals $\phi_{oa}(\vec r)$.

We have studied two different parameter sets for the inclusion of Mn impurities
into existing tight-binding models for GaAs. The first set
due to Ma{\v s}ek (model A) is derived from
a first principles approach while the second one due to
Tang and Flatt{\' e} (model B) is based on
phenomenologically deduced parameters.
In either case we checked our simulation by first calculating the
band structure and density of states for clean GaAs using the tetrahedron
method.\cite{Lehmann72} Then we considered disordered super-cells
which are repeated periodically, and their ensembles.
The sizes of the super-cells are typically between 128 and 1024 atomic sites
while the disorder averages were performed over 5 to 15 random configurations,
depending on the system size and the number of impurities.
The finiteness of the systems
limits the impurity concentrations to $x\gtrsim 0.006$.
As we will demonstrate below, the density of states
is self-averaging for super-cells of the
above-mentioned sizes so that increasing the number of
disorder averages decreases the fluctuations of the results but does not give
any new qualitative features. The self-averaging property also allows to imply
periodic boundary conditions to the super-cells by which less
fluctuating results can be obtained. These periodic repetitions
of the super-cells correspond to a summation over various $\vec k$ points
in the Brillouin zone. In our calculations we used between 20 and 505 different
points in $\vec k$ space which are chosen in close resemblance to the ones given by
the tetrahedron method. However, in contrast to the clean system, a
linear interpolation of the band structure does not give an
improvement of the quality of the results as the eigenenergies
of a disordered super-cell for a given $\vec k$ lie too close to each other.

\subsubsection{Model A (Ma{\v s}ek)}
The first model we study was
introduced by Ma{\v s}ek.\cite{Masek07a, Masek07b} It starts from a
tight-binding approach for the host material which
includes the first and second nearest neighbor interactions \cite{Talwar82a}
as well as spin orbit coupling.\cite{Masek07a, Chadi77a}
This gives a rather accurate description of the band structure
including the conduction bands away from the $\Gamma$ point.
The on-site and hopping terms for the Mn impurity in the
disordered system were
obtained from a first principles calculation using a self-consistent
Hartree-Fock-approximation for a single Mn
impurity hybridized in a GaAs lattice. The resulting parameters
were then checked within the same theoretical framework for a system with
10\% of the Ga atoms replaced by Mn and appeared to be robust.
This procedure gives for the on-site energies
$E_{\rm Mn,s\uparrow}=-0.4$~eV, $E_{\rm Mn,s\downarrow}=0.0$~eV, and
$E_{\rm Mn,p}=4.374$~eV. Additionally, the ten d-orbitals of Mn are
explicitly included. These orbitals can be divided into two
subgroups according to their symmetry, i.e., (${\rm Mn,t2}$) and (${\rm Mn,e}$).
The corresponding values are $E_{\rm Mn,t2\uparrow}=-2.21981$~eV,
$E_{\rm Mn,t2\downarrow}=2.63427$~eV, $E_{\rm Mn,e\uparrow}=-3.01348$~eV,
and $E_{\rm Mn,e\downarrow}=2.36445$~eV. The hybridization of these
d-levels with the s- and p-orbitals of the nearest neighbors is also
included. The relevant hopping parameters are $V_{sd,\sigma} = -1.1077$~eV,
$V_{pd,\sigma} = -1.0341$~eV, and $V_{pd,\pi} = 0.4767$~eV.
This model is supported by photoemission spectroscopy
experiments \cite{Okabayashi01} that revealed minor differences in the
band structure of ${\rm Ga}_{1-x}{\rm Mn}_x {\rm As}$ compared to GaAs
except for the appearance of the additional Mn d-states.

\subsubsection{Model B (Tang, Flatt{\'e})}
The second model we studied was introduced by Tang and
Flatt{\'e}.\cite{Tang04a,Kitchen06a}
The electronic structure of the GaAs host material is described
by a slightly different $sp^3$ tight-binding model which also includes
spin-orbit coupling but remains within
the nearest neighbor approximation.\cite{Chadi77a} Despite the
lack of the second nearest neighbor terms this model gives a fairly accurate
description of the band structure of GaAs around the $\Gamma$ point.
The replacement of a Ga atom by a Mn impurity is modelled by
an effective potential which describes the changes in on-site energies
and hybridization of the Mn d-orbitals with the As p-orbitals.
The Mn d-orbitals are included indirectly by a
spin-dependent effective energy shift of the nearest neighbor p-orbitals.
There are two independent parameters of this effective potential.
They are given by compensating the difference in the atomic ionization
energies and by fitting the experimentally \cite{Linnarsson97} obtained 
Mn acceptor level lying at $0.11$~eV.
This model was applied to describe single Mn atoms and Mn pairs in
GaAs and showed good agreement with experimental results for the
local and total density of states as well as the shape of
the wave functions.\cite{Kitchen06a}

%%%%%%%%%%%%%%%%%%%%%%%%%%%%%%%%%%%%%%%%%%%%
\section{Density of states}
\label{sec:density}
%%%%%%%%%%%%%%%%%%%%%%%%%%%%%%%%%%%%%%%%%%%%
\begin{figure}[t]
\resizebox{0.45\textwidth}{!}{\includegraphics{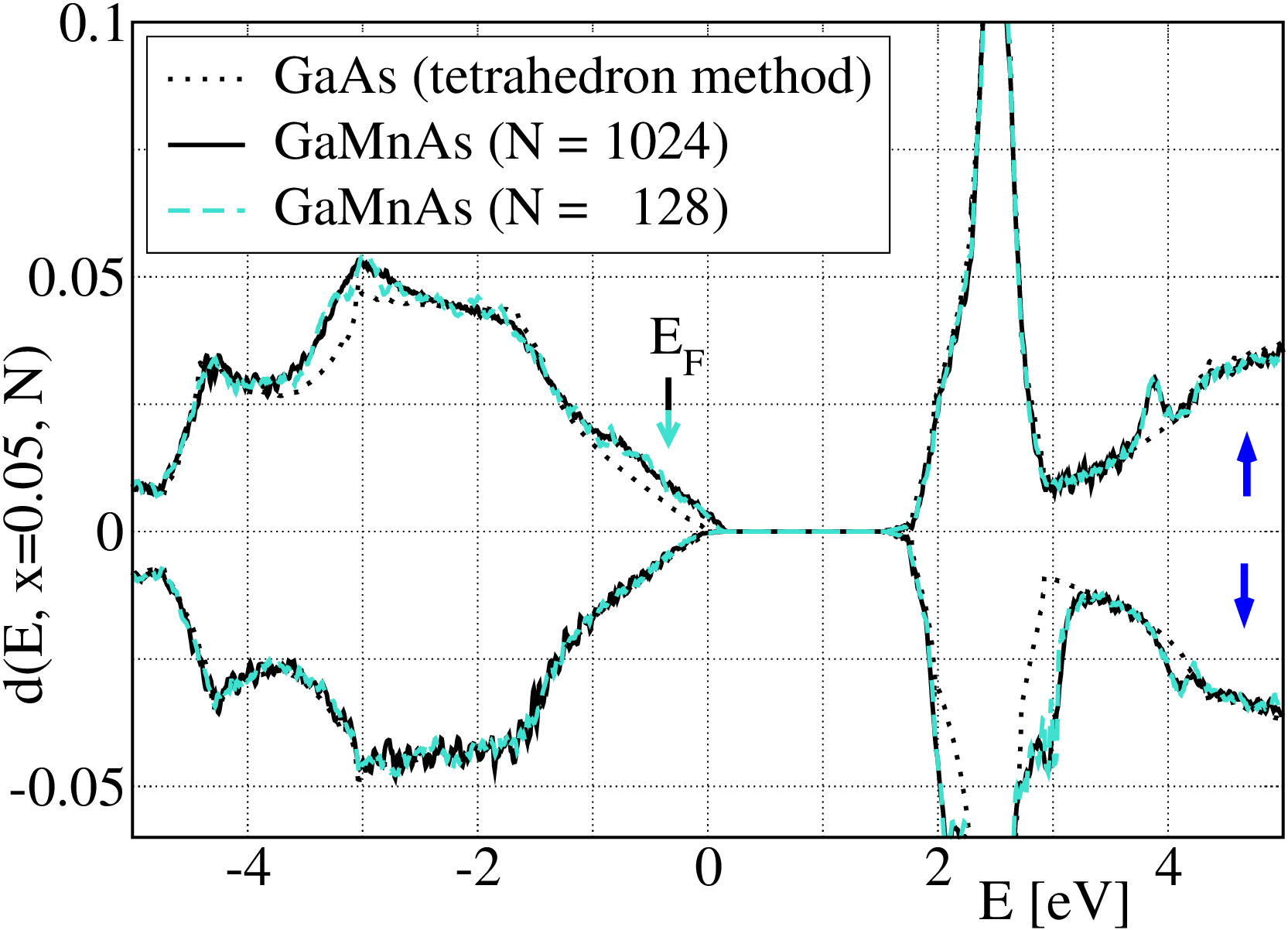}}
\caption{\label{fig:A_dos_size} 
 Model A -- Density of states for clean GaAs (black dotted, tetrahedron method)
 and disordered GaMnAs. The Mn
 concentration is $x=0.05$. System sizes are $N=128$ sites
 (505 $\vec k$ vectors, 15 disorder configurations) and
 $N=1024$ sites (20 $\vec k$ vectors, 5 disorder
 configurations). The arrow indicates the position of the Fermi energy,
 $E_{\rm F} = -0.34$~eV. The upper (lower) part of the
 figure shows the spin up (down) contribution to the density of states.
}
\end{figure}

\begin{figure}[t]
 \resizebox{0.45\textwidth}{!}{\includegraphics{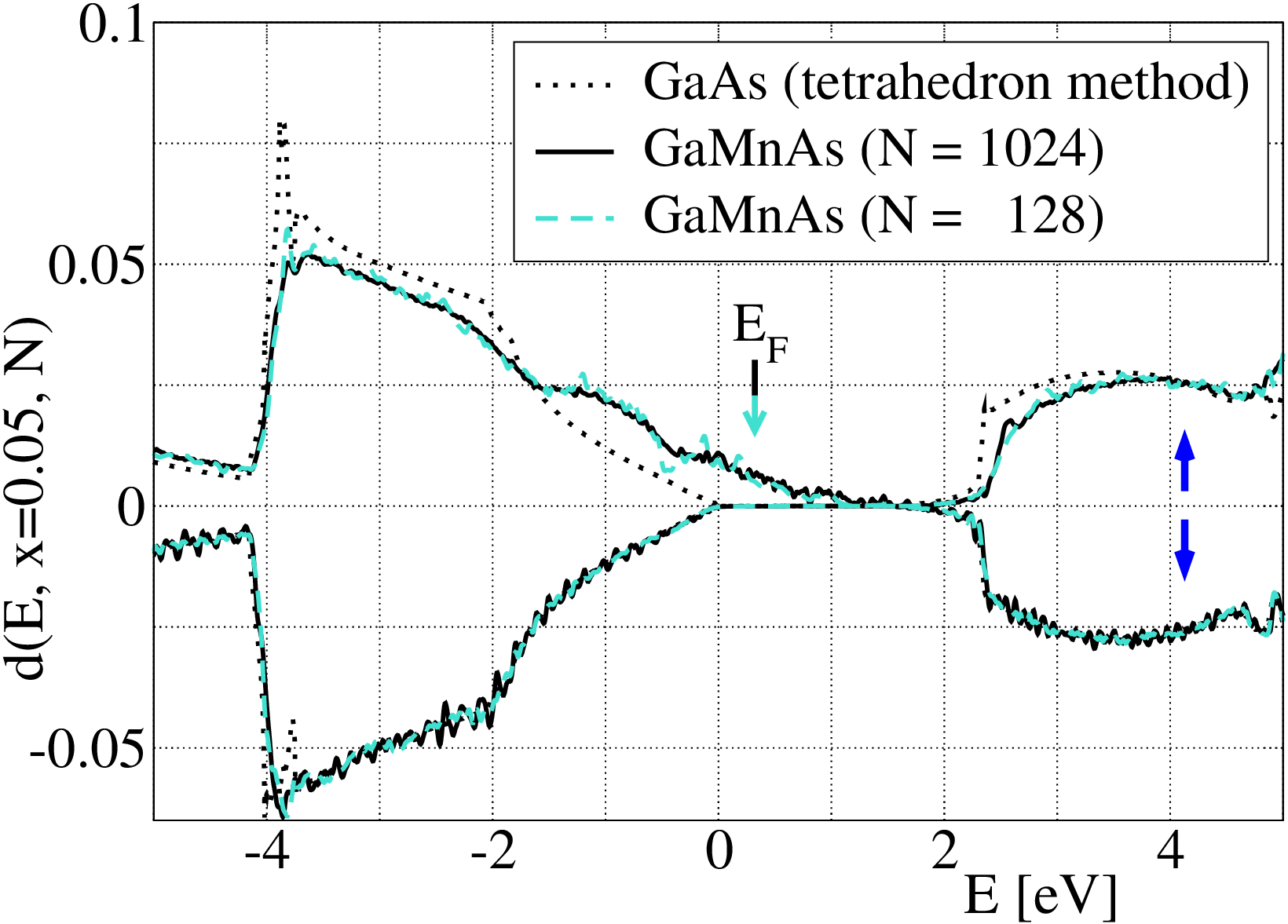}}
\caption{\label{fig:B_dos_size} 
 Model B -- Density of states for clean GaAs (black dotted, tetrahedron method)
 and disordered GaMnAs. The Mn
 concentration is set to $x=0.05$. System sizes are $N=128$ sites
 (505 $\vec k$ vectors, 15 disorder configurations) and
 $N=1024$ sites (20 $\vec k$ vectors, 5 disorder
 configurations). The arrow indicates the position of the Fermi energy,
 $E_{\rm F} = 0.30$~eV. The upper (lower) part of the
 figure shows the spin up (down) contribution to the density of states.
}
\end{figure}

The density of states, $d(E,x,N)$, is obtained by direct diagonalization of the
tight-binding Hamiltonian (\ref{eq:hamiltonian}) and the evaluation of the sum
\begin{equation}
 \label{eq:dos} d(E,x,N) = \frac{1}{n(\infty)} \, \sum\limits_{b, \vec k}
 \; \delta_\epsilon \left( E - E_{\vec k}^{(b)} \right) \, ,
\end{equation}
where $\delta_\epsilon(E)$ is a broadened delta-function of the chosen
width $\epsilon=5$ meV. The eigenvalues $E_{\vec k}^{(b)}$ of
the Hamiltonian are labeled
by the band index $b$. The density of states is normalized by the total
number of states $n(\infty )$. As the parameter sets for clean
GaAs are different for the two models the resulting densities
of states also show different behavior, in particular for
the conduction band, see Figs.~\ref{fig:A_dos_size} and \ref{fig:B_dos_size}.

The results for a Mn concentration $x=0.05$ and two different
system sizes are shown in Figs.~\ref{fig:A_dos_size} and \ref{fig:B_dos_size}
for the models A (Ma{\v s}ek) and B (Tang, Flatt{\'e}),
respectively. In both cases one can clearly observe
the self-averaging properties of the density of states --  the deviations
between the two systems of different sizes are on the same order as the fluctuations
due to the disorder. This self-averaging property of the density of states can be
observed for the entire range of concentrations studied. This implies that
the quantity $d(E,x,N) = d(E,x)$ is independent of the system size $N$
if the smaller fluctuations are neglected.

For model A (Ma{\v s}ek) one finds that the density of states is
only slightly perturbed in comparison to the density of states for the clean system.
This is in contrast to the results of model B (Tang, Flatt{\'e})
where, especially around the top of
the valence band of the host material, the number of spin-up states
is significantly enhanced. In either model the spin-down contribution of the
valence band remains unaltered.

\begin{figure}[t]
\resizebox{0.45\textwidth}{!}{\includegraphics{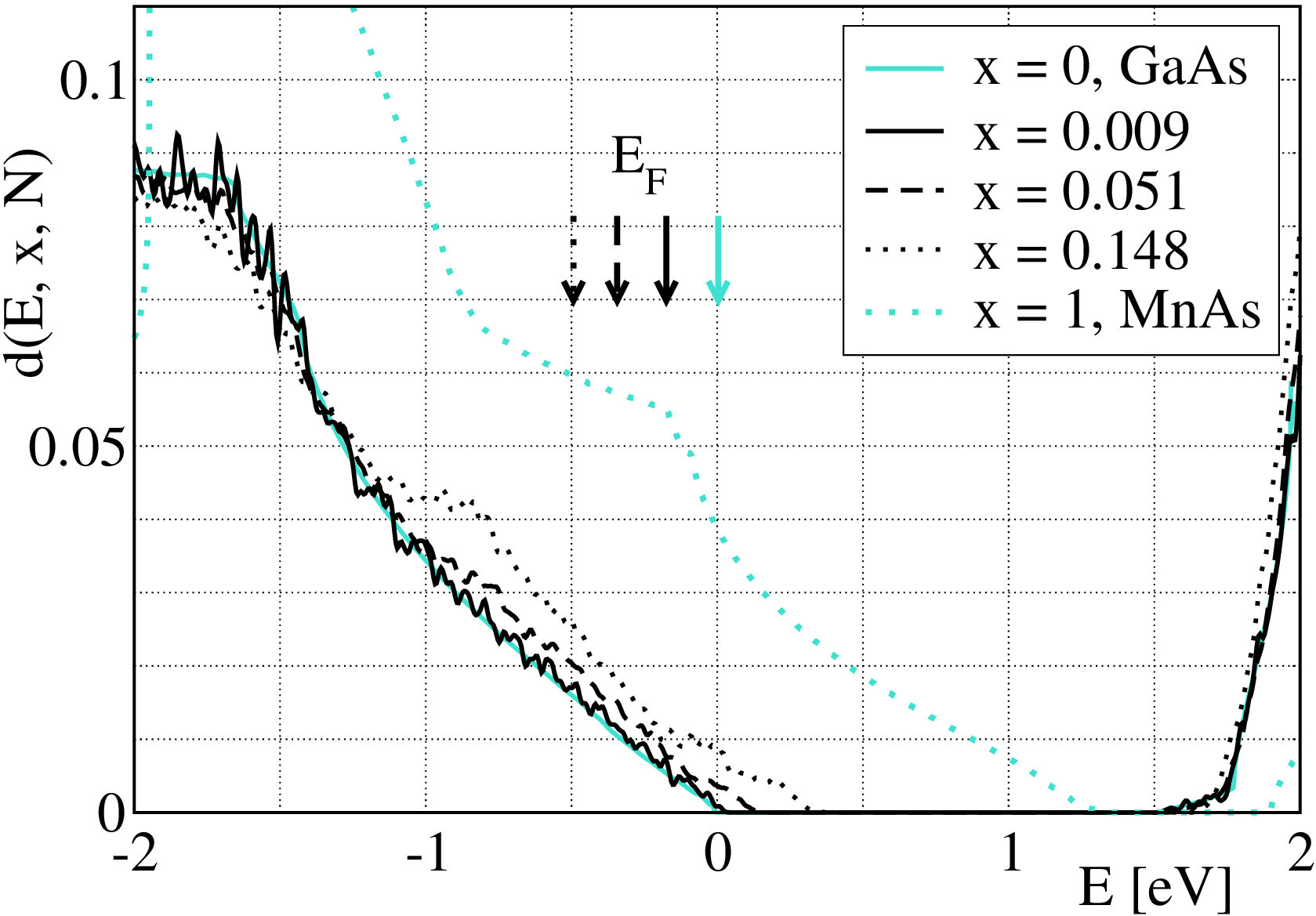}}
\caption{\label{fig:A_dos_conc} 
 Model A -- Density of states of GaMnAs. The Mn
 concentration is set to $x=0.009$ (1024 atoms,
 20 $\vec k$ vectors, 5 disorder configurations),
 $x=0.051$ (432 atoms,
 89 $\vec k$ vectors, 10 disorder configurations), and
 $x=0.148$ (432 atoms, 89 $\vec k$ vectors,
 10 disorder configurations). The arrows indicate the
 positions of the corresponding Fermi energies:
 $E_{\rm F}(x=0.009)=-0.17$~eV, $E_{\rm F}(x=0.051)=-0.34$~eV,
 $E_{\rm F}(x=0.148)=-0.49$~eV. For comparison the
 densities of states for clean GaAs and
 MnAs are shown. 
}
\end{figure}

\begin{figure}[t]
\resizebox{0.45\textwidth}{!}{\includegraphics{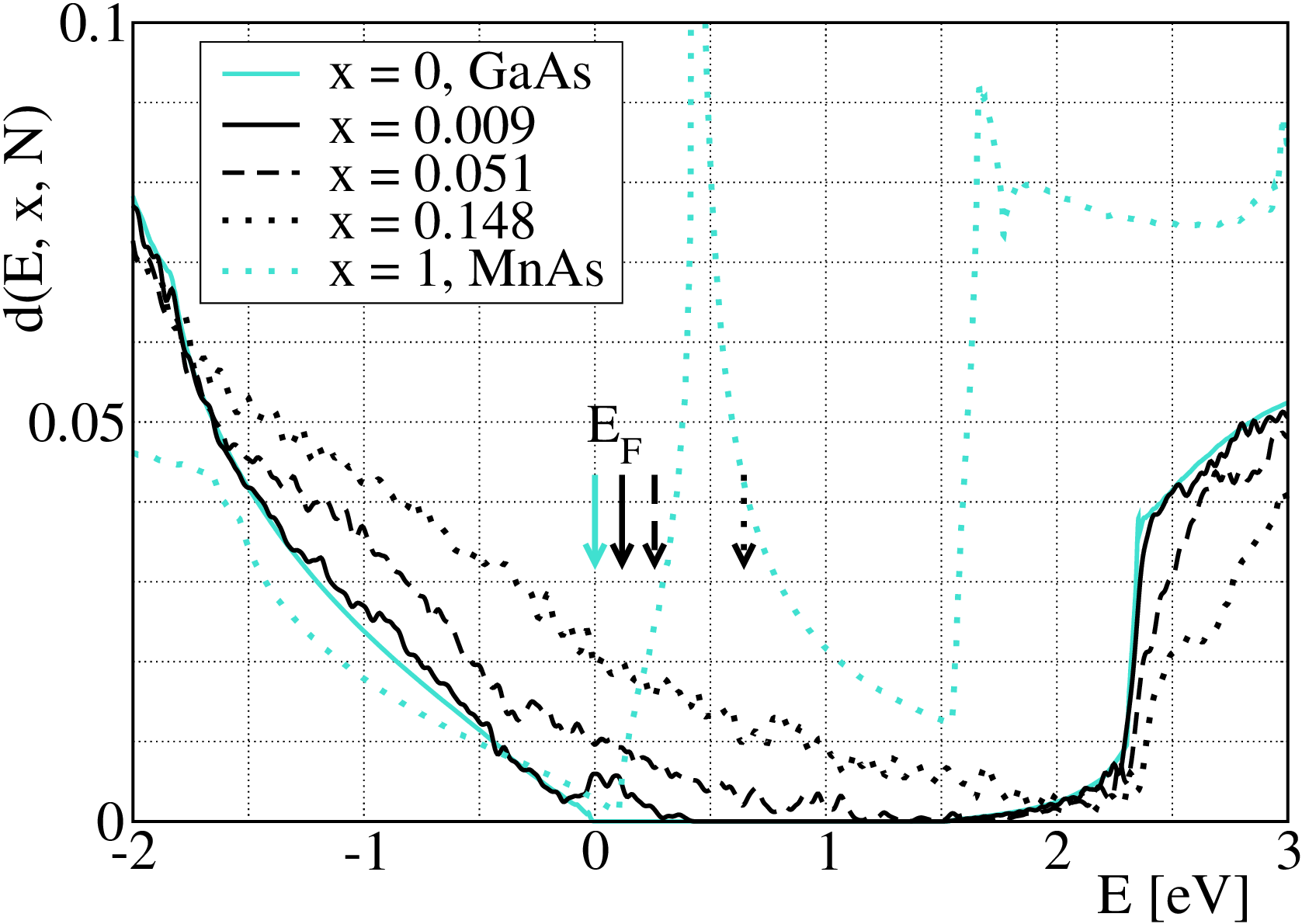}}
\caption{\label{fig:B_dos_conc}
 Model B -- Density of states of GaMnAs. The Mn
 concentration is set to $x=0.009$ (686 atoms), $x=0.051$ (432 atoms), and
 $x=0.148$ (432 atoms). In either case 89 $\vec k$ vectors and
 10 disorder configurations were used. The arrows indicate the
 positions of the corresponding Fermi energies:
 $E_{\rm F}(0.009)=0.10$~eV, $E_{\rm F}(0.051)=0.30$~eV,
 $E_{\rm F}(0.148)=0.65$~eV. For comparison the
 densities of states for clean GaAs and
 MnAs are shown.
 }
\end{figure}

\begin{figure}[t]
\resizebox{0.45\textwidth}{!}{\includegraphics{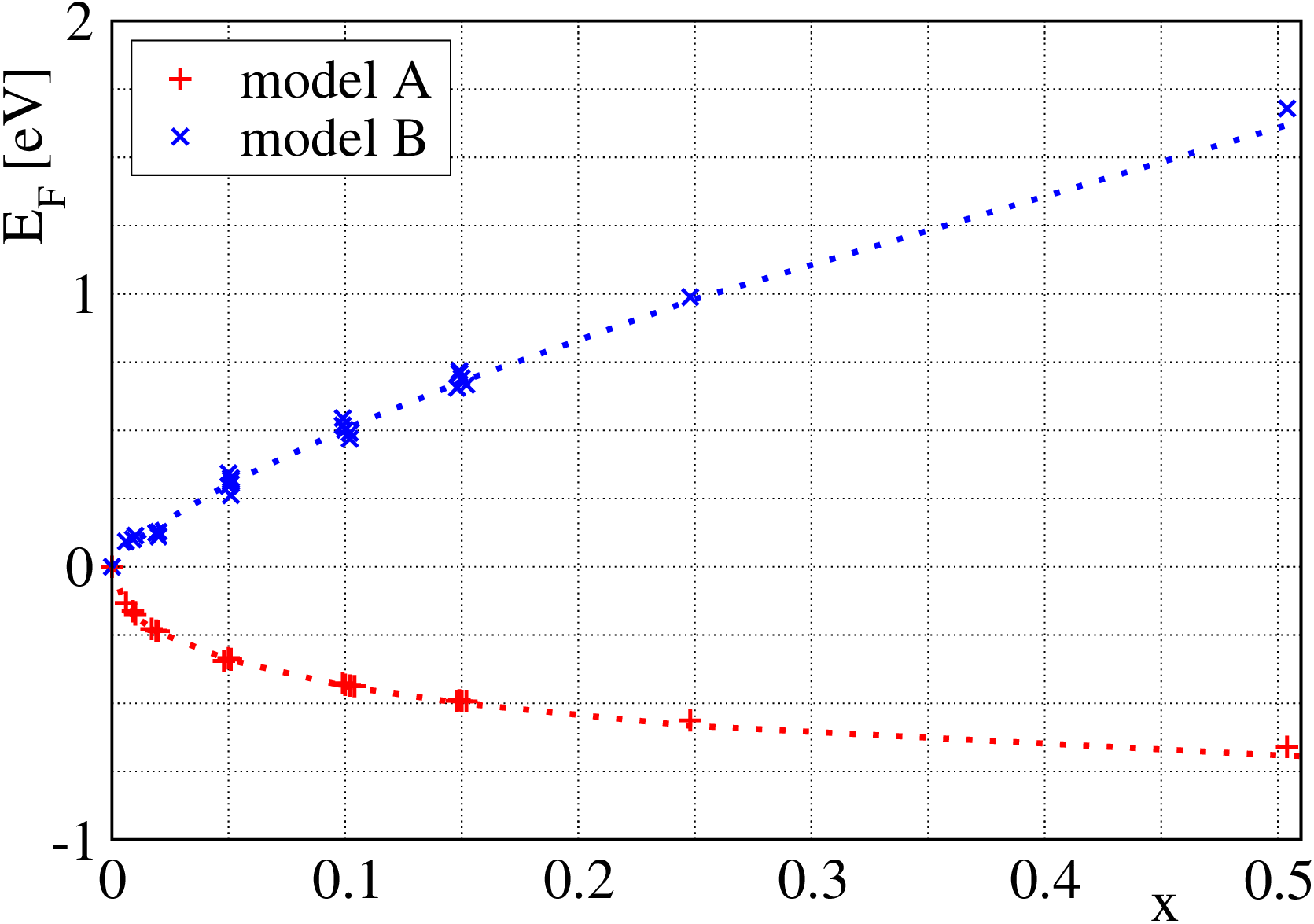}}
\caption{\label{fig:fermi}
 Model A, B -- Fermi energy of GaMnAs as a function of the impurity concentration
 measured from the top of the {\em GaAs host} valence band.
 The Fermi energy of model A moves into the host valence band
 while it increases into the host gap for model B. System
 sizes between 128 and 1024 atoms are shown. The dotted
 lines serve as a guide to the eye.
}
\end{figure}

\begin{figure}[t]
\resizebox{0.45\textwidth}{!}{\includegraphics{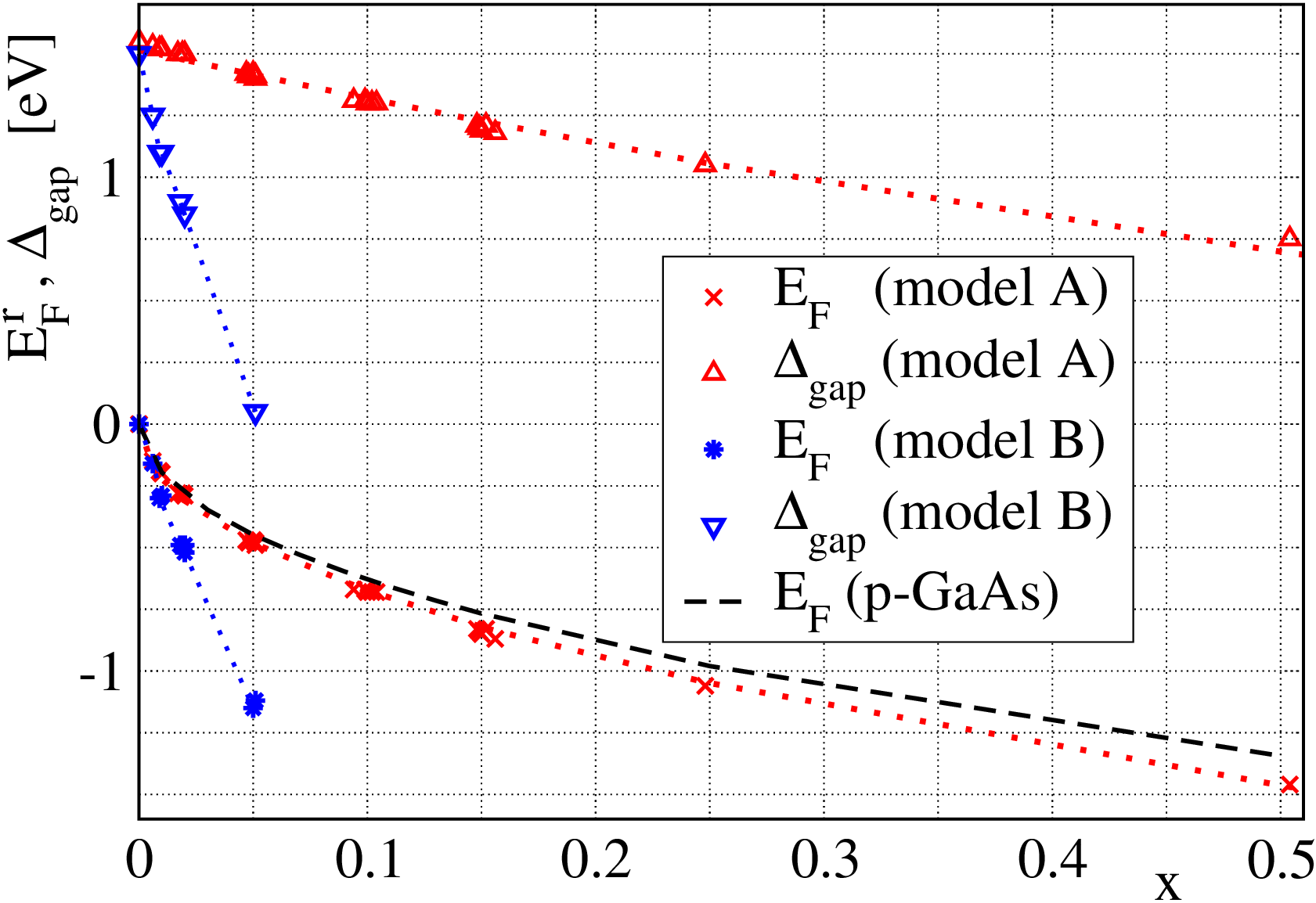}}
\caption{\label{fig:gap_fermi}
 Model A, B -- Fermi energy $E^{\rm r}_{\rm F}$
 measured relative to the top of the {\em GaMnAs} valence band and
 gap size $\Delta_{\rm gap}$.
 The dotted lines serve as a guide to the eye. The dashed line
 shows the Fermi energy of p-GaAs without impurities
 where the number of holes is adjusted
 to the corresponding GaMnAs impurity concentration $x$.
}
\end{figure}

As energies around the Fermi energy are most interesting
we present the density of states for a smaller energy window and
various Mn concentrations in the 
Figs.~\ref{fig:A_dos_conc} and \ref{fig:B_dos_conc}.
For model A (Ma{\v s}ek), increasing the Mn concentration from $1\%$
to $15\%$ does not change the
qualitative but only quantitative characteristics of the density of states.
In particular, no detached impurity
band forms within this model and the given range of Mn concentrations.
The size of the band gap decreases as some of the
spin-up states are shifted from below the top of the valence band into the gap
of the host material.
However, as adding Mn impurities also adds holes, the Fermi energy decreases and
moves deeper into the valence band,
see also Fig.~\ref{fig:fermi}.  Thus, the major
effect of an increasing number of Mn impurities is the larger number of holes
which leads to a lowering of the Fermi energy. This trend qualitatively follows
the Fermi energy of pure GaAs with the corresponding number of holes added,
see Fig.~\ref{fig:gap_fermi}.

For model B (Tang, Flatt{\'e}) a completely different
picture arises. Due to the strong effective potential of the Mn impurities there is
a large number of states shifted from the host valence band far into the gap.
Already at concentrations of $5\%$ there is almost no gap left at all.
The second effect of the addition of Mn impurities is again the
increased number of holes which lowers the Fermi energy. However,
for model B (Tang, Flatt{\'e}) this increasing number of holes
does not compensate the effect of the large number of states appearing
in the host gap. Hence one observes an increasing Fermi energy,
measured relative to the top of the host valence band, see Fig.~\ref{fig:fermi}.
The second major difference to model A (Ma{\v s}ek) is the appearance of a detached
impurity band for small enough impurity concentrations.
This impurity band starts to merge with the valence band for concentrations around $1\%$,
see Fig.~\ref{fig:B_dos_conc}, which is consistent with
experimental data.\cite{Blakemore73} Its center is positioned at the
impurity binding energy of Mn while its half width is on the order of $0.1$~eV.
For $x \gtrsim 0.05$ one cannot distinguish the
impurity band from the valence band anymore
as the two bands have merged completely.

The results for the Fermi energy and the band gap are summarized
in Figs.~\ref{fig:fermi} and \ref{fig:gap_fermi}. In Fig.~\ref{fig:fermi} the
Fermi energy of GaMnAs measured with respect to the top of the {\em GaAs host}
valence band is shown. From the behavior of the Fermi energy
as a function of the impurity concentration it
is evident that the two effects of adding
holes and shifting states are of unequal importance for the two models,
as described in the previous paragraphs.
In Fig.~\ref{fig:gap_fermi} we present the Fermi energy measured
with respect to the {\em GaMnAs} valence band. By choosing these
concentration dependent reference points for the energy we partially
suppress the effects due to the shifted states. Hence, the Fermi
energy decreases with respect to this energy reference point for
both models. Again one finds that the addition of the holes is the major effect
for model A (Ma{\v s}ek) as the resulting Fermi energy behaves very
similar to the case of clean p-GaAs with the corresponding number
of holes added, see Fig.~\ref{fig:gap_fermi}.
Furthermore we have included the results for the band gap
$\Delta_{\rm gap}$ into Fig.~\ref{fig:gap_fermi}. This information
can be used in the design and study of spintronic devices including
hole injection into GaMnAs. The
stronger influence of the disorder in model B (Tang, Flatt{\'e})
compared to model A (Ma{\v s}ek) becomes evident as the size
of the gap is much faster decreasing for model B. Experiments
based on scanning tunneling microscopy conductance spectra
resulted in an estimate $\Delta_{\rm gap} \approx 1.23$~eV for
$x\approx 0.03$ (Ref. \onlinecite{Tsuruoka02}) and, more recently,
photoconductivity measurements on GaMnAs heterostructures
yielded $\Delta_{\rm gap} \approx 1.41$~eV for
the same Mn concentration.\cite{Thomas07} The chemical potential
was found to be in the vincinity of the Mn impuritiy level.\cite{Thomas07}
However, the considered samples were unannealed or annealed only for
short times implying a significantly reduced carrier concentration
due to Mn intersticials. Therefore, a direct comparison with our data
is not possible and further, more detailed, experimental investigations
are necessary in order to decide which model provides a more accurate
description.

The differences for the density of states
between the two models can be better understood when the
density for a clean MnAs ($x=1$) system is calculated.
As zinc-blend MnAs does not form a stable configuration \cite{Sanvito00a}
this calculation does not correspond to a real physical system. Nevertheless
it describes the $x\to 1$ limit of the two models giving a good indication
of what to expect for increasing $x$.
The parameterization of model A (Ma{\v s}ek) results in an increased number
of states for MnAs in the gap of GaAs. For the
disordered GaMnAs there is a clear tendency for the density of states
to change from the GaAs-shape ($x=0$) towards the MnAs shape ($x=1$).
This can, for example, be seen by the formation of the shoulder in the
density of states at $E\approx -1$~eV and $x=0.148$,
see Fig.~\ref{fig:A_dos_conc}. For model B (Tang, Flatt{\'e}), the MnAs
density shows a qualitatively new feature at $E\approx 0.5$~eV
in the gap of the GaAs, seen in Fig.~\ref{fig:B_dos_conc}.
The appearing peak is due to the shift in the
energies of the p-states which is reflected in the rapidly vanishing gap for
the disordered GaMnAs.

%%%%%%%%%%%%%%%%%%%%%%%%%%%%%%%%%%%%%%%%%%%%
\section{Inverse participation ratio}
\label{sec:inverse}
%%%%%%%%%%%%%%%%%%%%%%%%%%%%%%%%%%%%%%%%%%%%
Besides the density of states and 
the Fermi energy we also studied the character of the eigenstates, in particular
their localization properties. To this end we analyzed the inverse participation
ratio averaged over all states in a given energy window:
\begin{equation}
 \label{eq:ipr} I(E, x, N) = \frac{ \sum\limits_{b, \vec k, a}
 \left| \sum\limits_o \left| c_{\vec k o a}  \right|^2 \right|^2
 \delta_\epsilon \left( E - E_{\vec k}^{(b)} \right)}{ \sum\limits_{b, \vec k}
 \delta_\epsilon \left( E - E_{\vec k}^{(b)} \right)}.
\end{equation}
Qualitatively, the participation ratio corresponds to the average number $N_{\rm occ}$
of sites occupied by the eigenstates of given energy $E$, i.e. $1/I \sim N_{\rm occ}$.
Thus, $I(E, x, N)$ scales with the inverse system size for extended states
as the probability $\sum_o | c_{\vec k o a} |^2$ to find the
electron at a certain atom is on the order of the inverse system size.
For strongly localized states $I(E, x, N)$ remains constant
independently of the system size.

\begin{figure}[t]
\resizebox{0.45\textwidth}{!}{\includegraphics{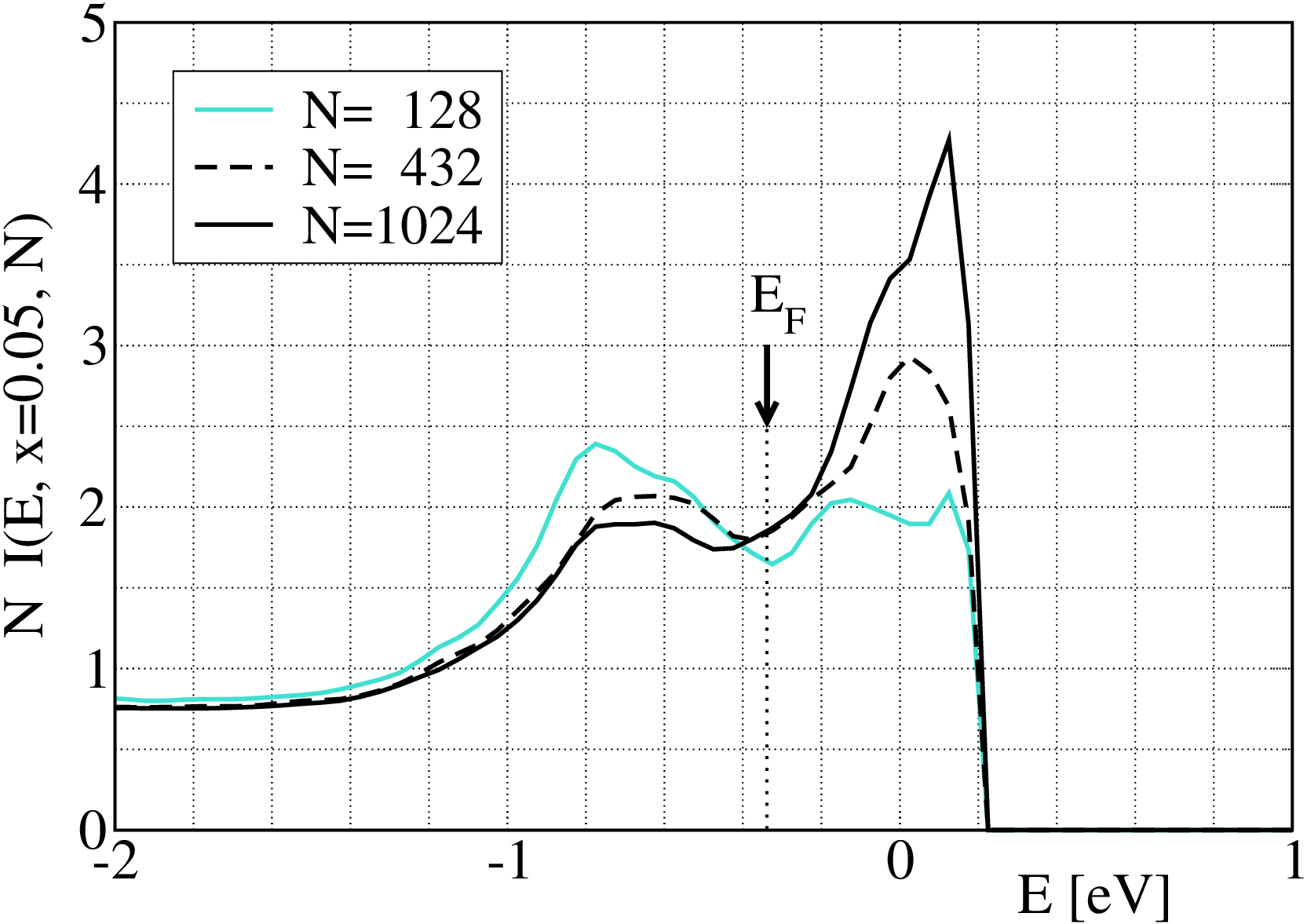}}
\caption{\label{fig:A_ipr_size}
 Model A -- Scaled inverse participation ratio $[N \cdot I(E,x,N)]$ for fixed
 concentration $x=0.05$ and various system sizes
 $N=128$ (505 $\vec k$ vectors, 15 configurations),
 $N=432$ (89 $\vec k$ vectors, 10 configurations), and $N=1024$
 (20 $\vec k$ vectors, 5 configurations). The arrow indicates the
 position of the Fermi energy $E_{\rm F} = -0.34$ eV.
 }
\end{figure}

\begin{figure}[t]
\resizebox{0.45\textwidth}{!}{\includegraphics{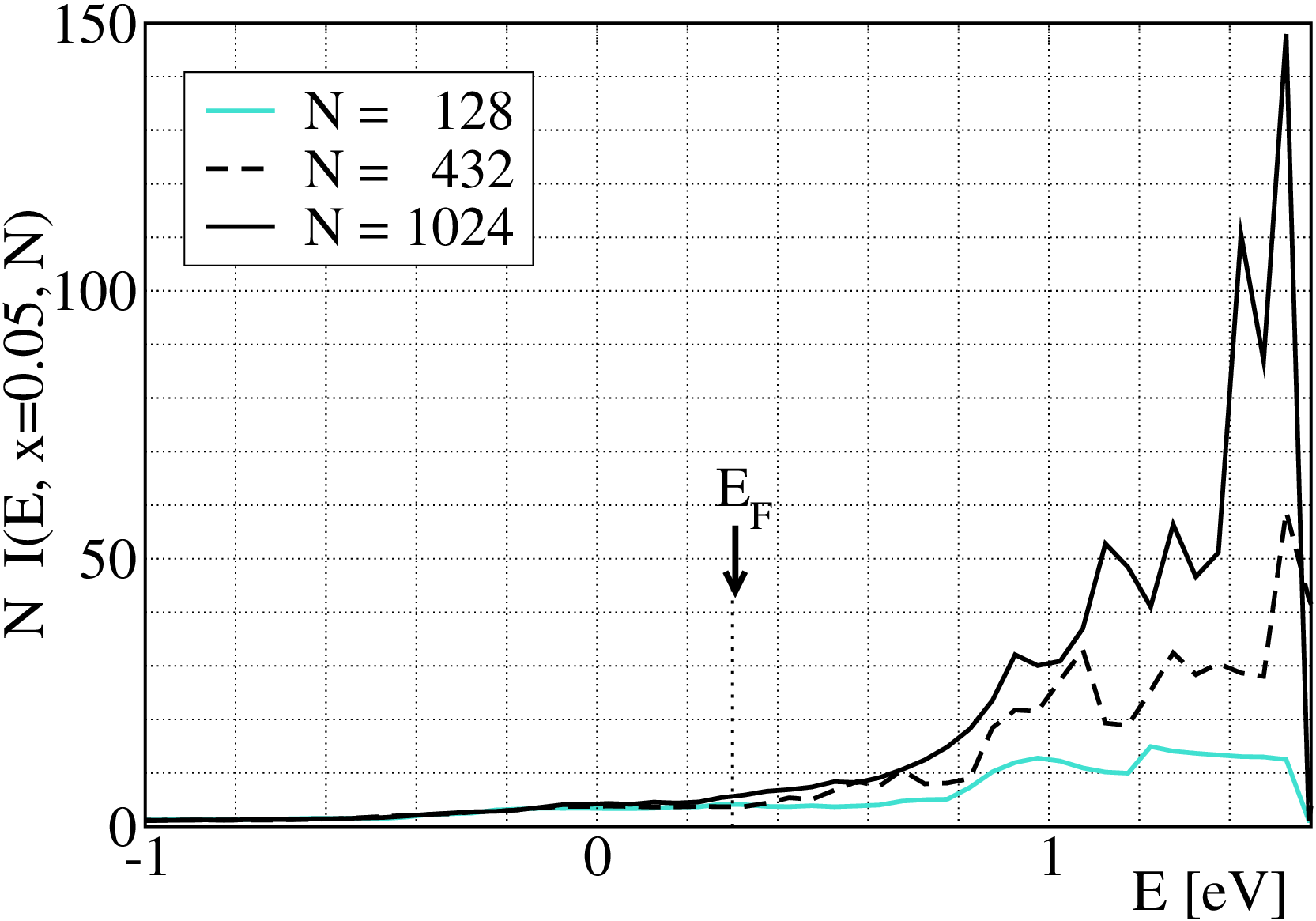}}
\caption{\label{fig:B_ipr_size}
 Model B -- Scaled inverse participation ratio $[N \cdot I(E,x,N)]$ for fixed
 concentration $x=0.05$ and various system sizes
 $N=128$ (505 $\vec k$ vectors, 15 configurations),
 $N=432$ (89 $\vec k$ vectors, 10 configurations), and $N=1024$
 (20 $\vec k$ vectors, 5 configurations). The arrow indicates the
 position of the Fermi energy $E_{\rm F} = 0.30$ eV.
 }
\end{figure}

\begin{figure}[t]
 \resizebox{0.45\textwidth}{!}{\includegraphics{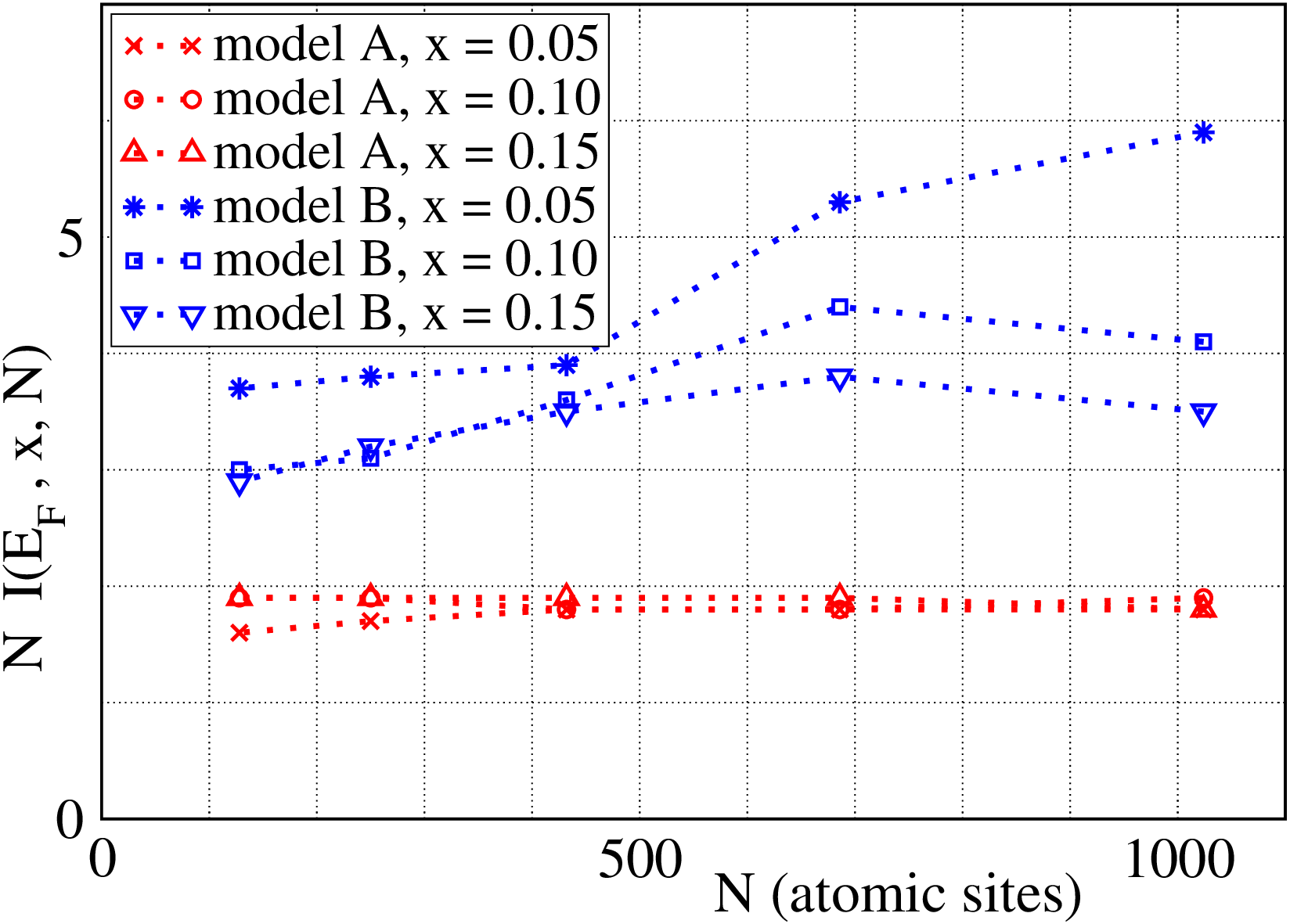}}
\caption{\label{fig:ipr_vs_size_conc}
 Model A, B -- Scaled inverse participation ratio
 $[N \cdot I(E_{\rm F},x,N)]$ at the Fermi energy
 as a function of the system size $N$. Mn concentrations
 $x=0.05$, $0.10$, and $0.15$
 are shown for model A and B.
}
\end{figure}

First, we consider the scaling of $[N\cdot I(E,x,N)]$ with system size $N$ for
a fixed impurity concentration. The results for the two models
are shown in Figs.~\ref{fig:A_ipr_size} and \ref{fig:B_ipr_size} where
$[N \cdot I(E, x=0.05, N)]$ is plotted.
As expected one finds extended states for the energies deep within the valence
band. As the energy is increased towards the Fermi energy, the value of
$[N \cdot I(E, x, N)]$ increases, meaning that on average 
less atoms participate in the corresponding states.  Comparing
Figs.~\ref{fig:A_ipr_size} and \ref{fig:B_ipr_size} one finds this effect to
be much stronger for model B (Tang, Flatt{\'e}). Nevertheless a clear linear
scaling of $[N \cdot I(E_{\rm F}, x, N)]$ with system size $N$, which would
indicate strongly localized states at the Fermi energy, cannot be found, see
Fig.~\ref{fig:ipr_vs_size_conc}. This implies that the
states around the Fermi energy reside on a subset of lattice
atoms; yet they are still extended throughout the sample. This holds true
for concentrations between $5\%$ and $15\%$, see
Fig.~\ref{fig:ipr_vs_size_conc}.
For model B (Tang, Flatt{\'e}), this effect is more pronounced and we will
argue that the states are spread over the sub-lattice of Mn sites.
As the density of states
decreases for energies closer to the band edge there are less
states giving a contribution to the inverse participation ratio
(\ref{eq:ipr}). This causes the stronger fluctuations in $I(E, x, N)$
that can be observed for energies around and above the Fermi energy.

Going from the Fermi energy $E_{\rm F}$
to higher energies close to the band edge, $E_{\rm be}$, the states
remain delocalized for model A (Ma{\v s}ek) while they show strong localization
for model B (Tang, Flatt{\'e}). Qualitatively this conclusion is supported
by the following argument. If an impurity state is strongly localized
it cannot overlap with other impurity states located at neighboring
impurity sites. This implies that the number of sites occupied by
an impurity state must be significantly smaller than $N/N_{\rm Mn}$.
As the participation ratio is a measure for the number
of occupied sites $N_{\rm occ}$ it follows that the states are strongly
localized only if $I(E,x,N)$ is large enough so that
$x \lesssim I(E, x, N)$ independent of the system size $N$.
This is consistent with the
statement that $I(E, x, N)$ scales inversely with the system size
$N$ for extended states while it remains constant for localized states.
For model A (Ma{\v s}ek) the condition $x \lesssim I(E, x, N)$ is never fullfilled in the
entire energy range, see Fig.~\ref{fig:A_ipr_size}.
However, for model B (Tang, Flatt{\'e}), going beyond the Fermi energy
this criterion can be fullfilled implying that there
are localized states, see Fig.~\ref{fig:B_ipr_size}.

\begin{figure}[t]
 \resizebox{0.45\textwidth}{!}{\includegraphics{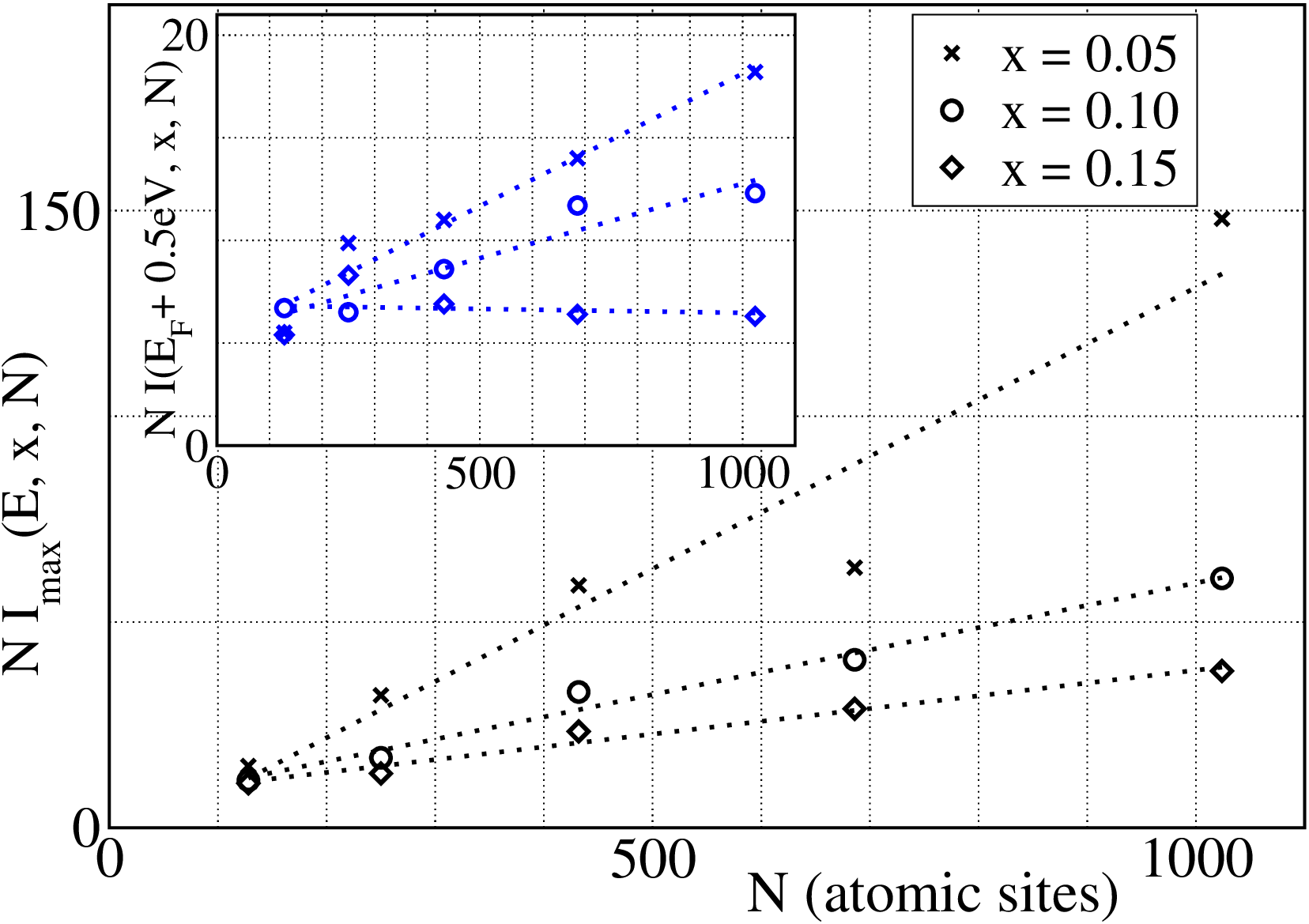}}
\caption{\label{fig:B_iprmax_size}
 Model B -- Maximum value of the scaled 
 inverse participation ratio $[N \cdot I_{\rm max}(E_{\rm be},x,N)]$ 
 for energies close to the valence band edge
 as a function of the system size $N$. The inset shows the same quantity
 at energies $E_{\rm F} + 0.5$ eV.
}
\end{figure}

To demonstrate this localization property for model B (Tang, Flatt{\'e})
in more detail
we show the maximum values of $[N \cdot I(E, x, N)]$, which correspond
to energies $E_{\rm be}$ close to the band edge,
in Fig.~\ref{fig:B_iprmax_size}.
The clear linear increase of the scaled inverse participation
ratio for the states closer to the band edge indicates
strong localization at $E$ around $E_{\rm be}$. On the other hand,
at lower energies $E=E_{\rm F} + 0.5$~eV
a transition from delocalized to localized states takes place depending
on the concentration of Mn impurities, see the inset of
Fig.~\ref{fig:B_iprmax_size}. The
linear increase of $[N \cdot I(E_{\rm F} + 0.5\;{\rm eV}, x, N)]$
with size $N$ indicates localized states for concentrations
$x<0.15$. On the other hand, the scaled inverse participation ratio remains
constant for $x=0.15$ for all system sizes meaning that the states are extended.
Obviously, the concentration of Mn impurities is in this case large enough
for the impurity states to overlap.
Generally one can assume that for each fixed energy $E$ there is a critical
concentration $x_{\rm crit}(E)$ for which the slope of
$[N \cdot I(E, x, N)]$ vanishes. Based on this critical concentration one
can estimate the spatial extension of the Mn impurity states. The
average distance of impurities for given concentration
$x$ is approximately $0.4\;{\rm nm} \cdot x^{-1/3}$. For the critical 
concentration $x_{\rm crit}(E)$ the impurity states start to overlap
as the transition from localization to delocalization takes place.
Therefore the spatial extension of the states at a fixed energy can
be estimated by $0.4\;{\rm nm} \cdot x^{-1/3}_{\rm crit}(E)$.
From Fig.~\ref{fig:B_ipr_size} we find
$x_{\rm crit}(E_{\rm F} + 0.5{\rm eV}) \approx 0.15$
implying that the impurity states extend over a range of
$\approx 0.7$~nm. At higher energies, $E_{\rm be}$,
extrapolation of our
data leads to $x_{\rm crit}(E_{\rm be}) \approx 0.18$ and
a spatial extension of the impurity states of $\approx 0.6$~nm.
This estimate is consistent with the experimentally deduced
value.\cite{Nagai01}

\begin{figure}[t]
\resizebox{0.45\textwidth}{!}{\includegraphics{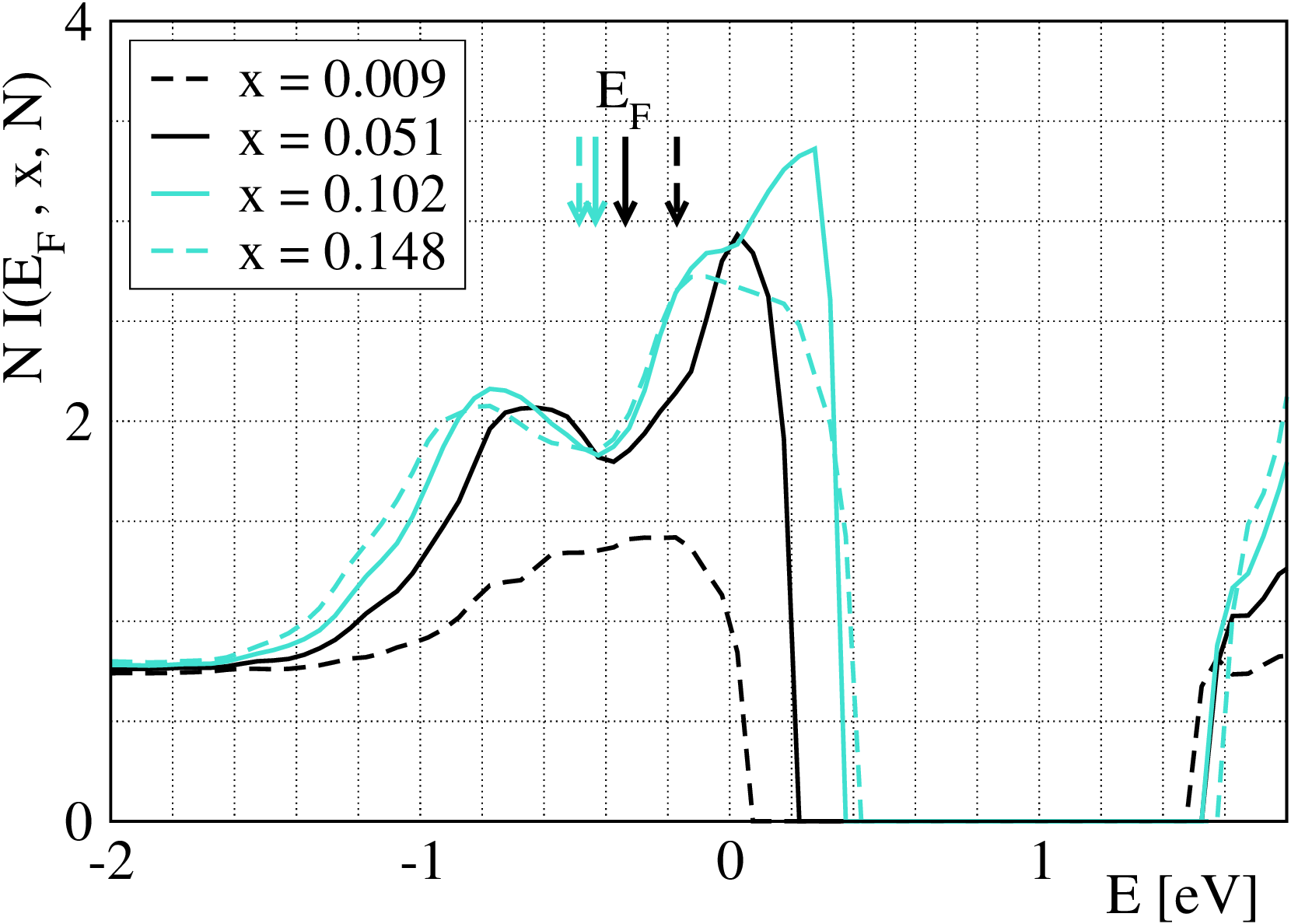}}
\caption{\label{fig:A_ipr_conc}
 Model A -- Scaled inverse participation ratio $[N \cdot I(E,x,N)]$ for
 various concentrations $x=0.009$ (1024 atoms, 20 $\vec k$ vectors, 5 disorder
 configurations), $x=0.051$, $0.102$, and $0.148$ (432 atoms, 89 $\vec k$ vectors, 10
 disorder configurations). The arrows indicate the
 positions of the Fermi energy.
}
\end{figure}

\begin{figure}[t]
 \resizebox{0.45\textwidth}{!}{\includegraphics{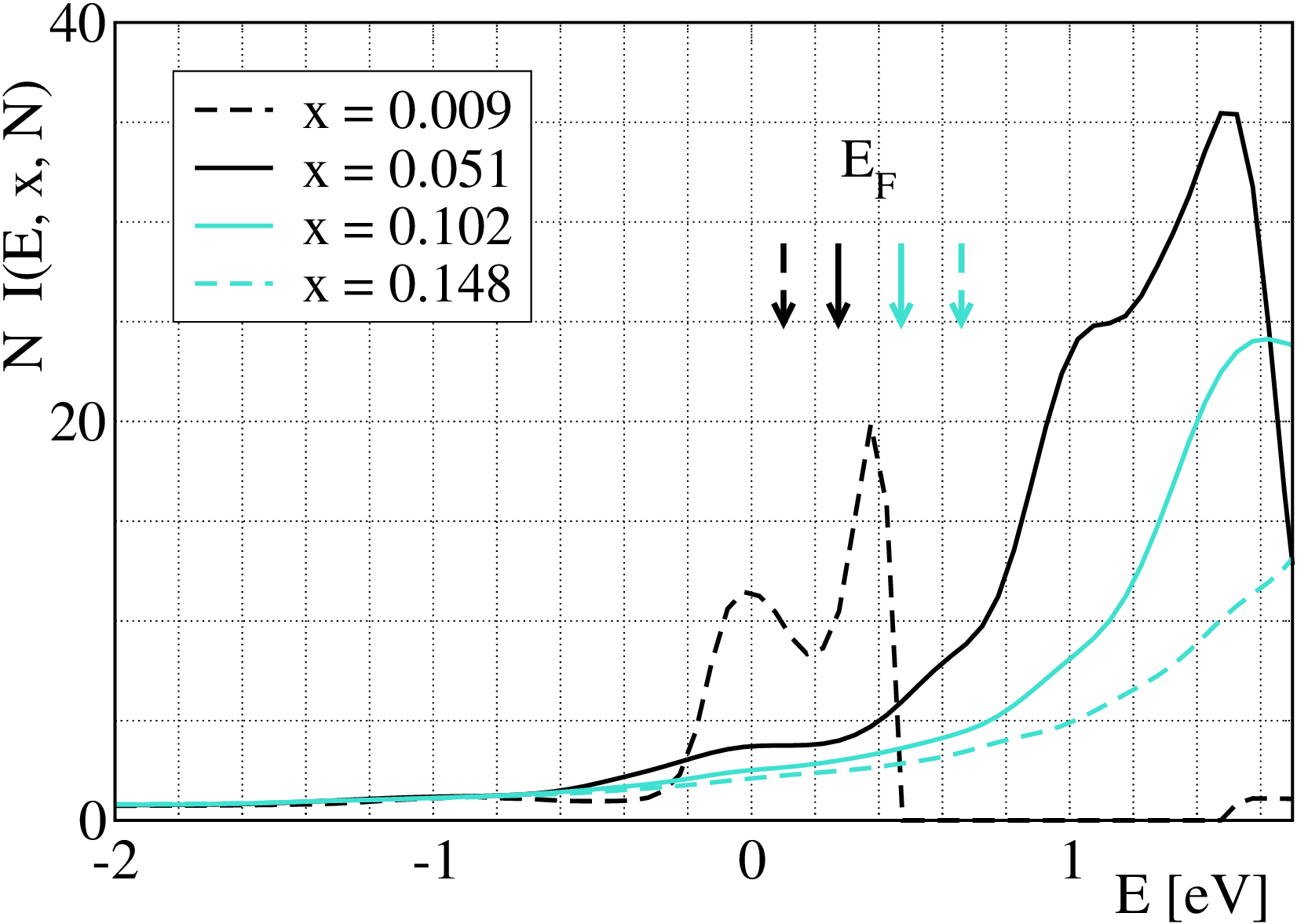}}
\caption{\label{fig:B_ipr_conc}
 Model B -- Scaled inverse participation ratio $[N \cdot I(E,x,N)]$ for
 various concentrations $x=0.009$ (1024 atoms, 20 $\vec k$ vectors, 5 disorder
 configurations), $x=0.051$, $0.102$, and $0.148$ (432 atoms, 89 $\vec k$ vectors, 10
 disorder configurations). The arrows indicate the
 positions of the Fermi energy.
}
\end{figure}

\begin{figure}[t]
 \resizebox{0.45\textwidth}{!}{\includegraphics{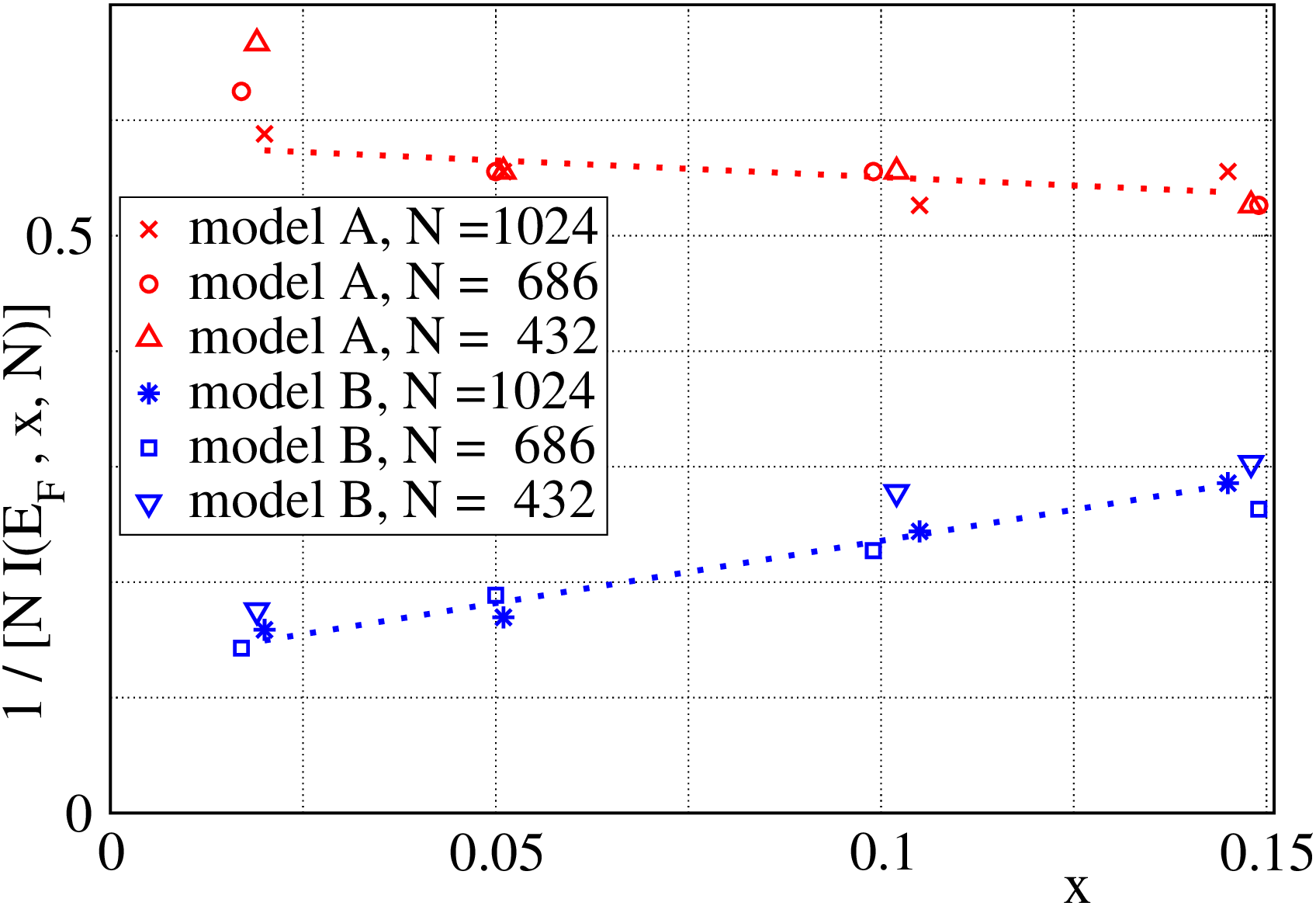}}
\caption{\label{fig:ipr_vs_conc_size}
 Model A, B -- Scaled participation ratio
 $1/[N \cdot I(E_{\rm F}, x, N)]$ at the Fermi energy
 as a function of the Mn concentration $x$. System sizes of
 $N=1024$, $686$, and $432$
 atoms are shown for model A and B.
}
\end{figure}

In a second step we analyze the influence of the Mn concentration
on the inverse participation ratio (\ref{eq:ipr}). Figures~\ref{fig:A_ipr_conc}
and \ref{fig:B_ipr_conc} show the results for a fixed system size $N=432$ for
models A (Ma{\v s}ek) and B (Tang, Flatt{\'e}), respectively.
For comparison, we have also included the
result of a larger system with a smaller concentration of Mn impurities.
In the case of model A (Ma{\v s}ek), the increase of $[N\cdot I(E,x,N)]$
with increasing energy starts first for the highest concentration $x=0.15$.
The reason can be found in the density of states -- the larger $x$ is, the stronger
the valence band changes. However, if one considers $[N\cdot I(E_{\rm F},x,N)]$
at the Fermi energy, which also depends on $x$, the influence
of the concentration is negligible, see Fig.~\ref{fig:ipr_vs_conc_size},
where $1/[N \cdot I(E_{\rm F}, x, N)]$ is plotted as a function of $x$.

For model B (Tang, Flatt{\'e}) the situation is
different since $[N\cdot I(E_{\rm F},x,N)]$ increases with
decreasing concentration $x$, as shown in Fig.~\ref{fig:B_ipr_conc},
meaning that the number of occupied sites increases with $x$.
This suggests that the states around the
Fermi energy tend to the Mn sub-lattice. This interpretation is
supported by the fact that $1/[N\cdot I(E_f,x,N)]$ increases 
linearly with the concentration $x$, see Fig.~\ref{fig:ipr_vs_conc_size}.
As $1/[N\cdot I(E_f,x,N)]$ is a measure for the relative number of atoms where the
states reside one can conclude that for model B (Tang, Flatt{\'e})
the states tend to spread
on the Mn sub-lattice and nearby sites. As an independent
check of this conclusion we further looked at the probability of states
with given energy to be found on the Mn impurities.
As a result (not presented here) we found that this probability
is indeed strongly increased already at the Fermi energy. It increases to almost
one at even higher energies $E_{\rm be}$ close to the band edge.

%%%%%%%%%%%%%%%%%%%%%%%%%%%%%%%%%%%%%%%%%%%%
\section{Conductivity}
\label{sec:conductivity}
%%%%%%%%%%%%%%%%%%%%%%%%%%%%%%%%%%%%%%%%%%%%
The optical conductivity is a directly measurable quantity which is
closely related to the absorption. As there have been various
experiments studying this quantity for GaMnAs we analyze
the conductivity obtainable from the two models A (Ma{\v s}ek)
and B (Tang, Flatt{\'e}) and compare our results with
the experimental findings. The starting point for our
analysis is the expression for the real part of the
conductivity,
\begin{equation}
 \label{eq:conductivity}
 {\rm Re} \, \sigma(\Delta) = \frac{\pi e^2 \hbar}{m^2 \Delta}
 \sum\limits_{i,f} \int\limits_{\rm BZ} \frac{{\rm d}^3k}{(2\pi)^3}
 \left|\langle \Psi_{\vec k}^{(f)}|p_z|\Psi_{\vec k}^{(i)} \rangle \right|^2
 \, \delta(E_{\vec k}^{(f)} - E_{\vec k}^{(i)} - \Delta) \, ,
\end{equation}
where $f$ ($i$) labels the final (initial) states and
$p_z$ is the $z$-component of the momentum operator.
We restrict our considerations to a linearly polarized field in z-direction.
As we investigate the optical conductivity in the infrared regime
the eigenstates must share the same wavevector $\vec k$.
The resulting conductivity cannot be expected to be
quantitatively exact as tight-binding approaches are known to
underestimate the absorption.\cite{Ren81,Jancu98}
Nevertheless, the order of magnitude of various related quantities, like
effective masses, and important qualitative
conclusions can be extracted. Within the tight-binding
approach the matrix element of the momentum operator
in Eq.~\ref{eq:conductivity} can be expressed
in terms of the Hamiltonian matrix, Eq.~\ref{eq:hamiltonian},
and the distance between the localized orbitals.\cite{Theodorou99}
This approximation neglects the spin-orbit coupling
for the evaluation of the matrix element itself but 
correlations of the density of states are treated correctly giving
a reasonable result for the optical conductivity.\cite{Voon93}

\begin{figure}[t]
 \resizebox{0.45\textwidth}{!}{\includegraphics{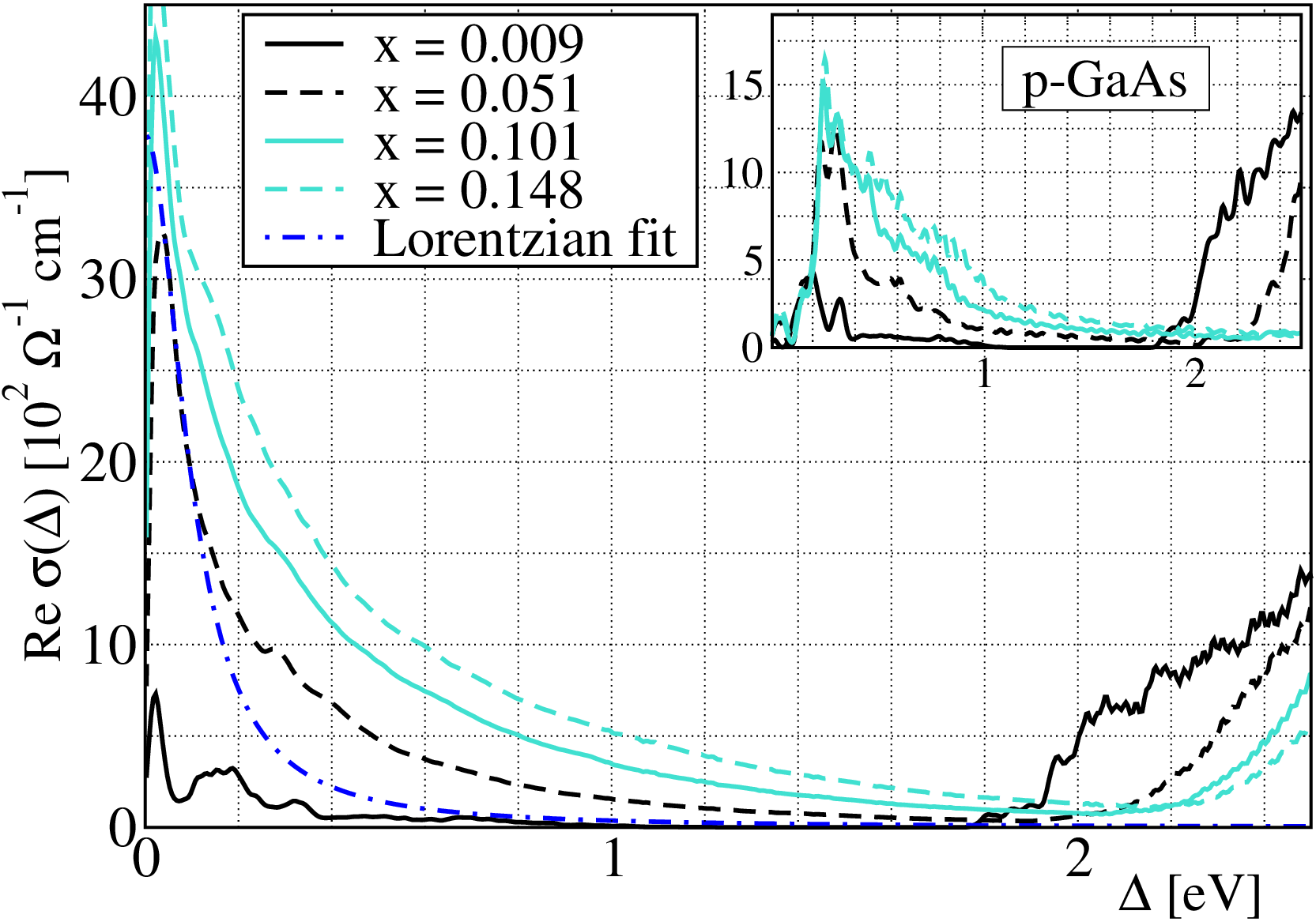}}
\caption{\label{fig:A_abs_conc}
 Model A -- Conductivity ${\rm Re} \, \sigma(\Delta,x)$ for
 various concentrations $x=0.009$ (686 atoms, 89 $\vec k$ vectors)
 and $x=0.051$, $0.101$, and $0.148$ (432 atoms, 240 $\vec k$ vectors).
 The results are averaged over 8 different disorder configurations.
 The inset shows the results for p-doped GaAs
 in the clean limit of the model with the corresponding number of holes
 added. The dashed-dotted line is a Lorentzian fit to the Drude peak
 at $\Delta = 0$ for the $x=0.051$ data.
}
\end{figure}

\begin{figure}[t]
 \resizebox{0.45\textwidth}{!}{\includegraphics{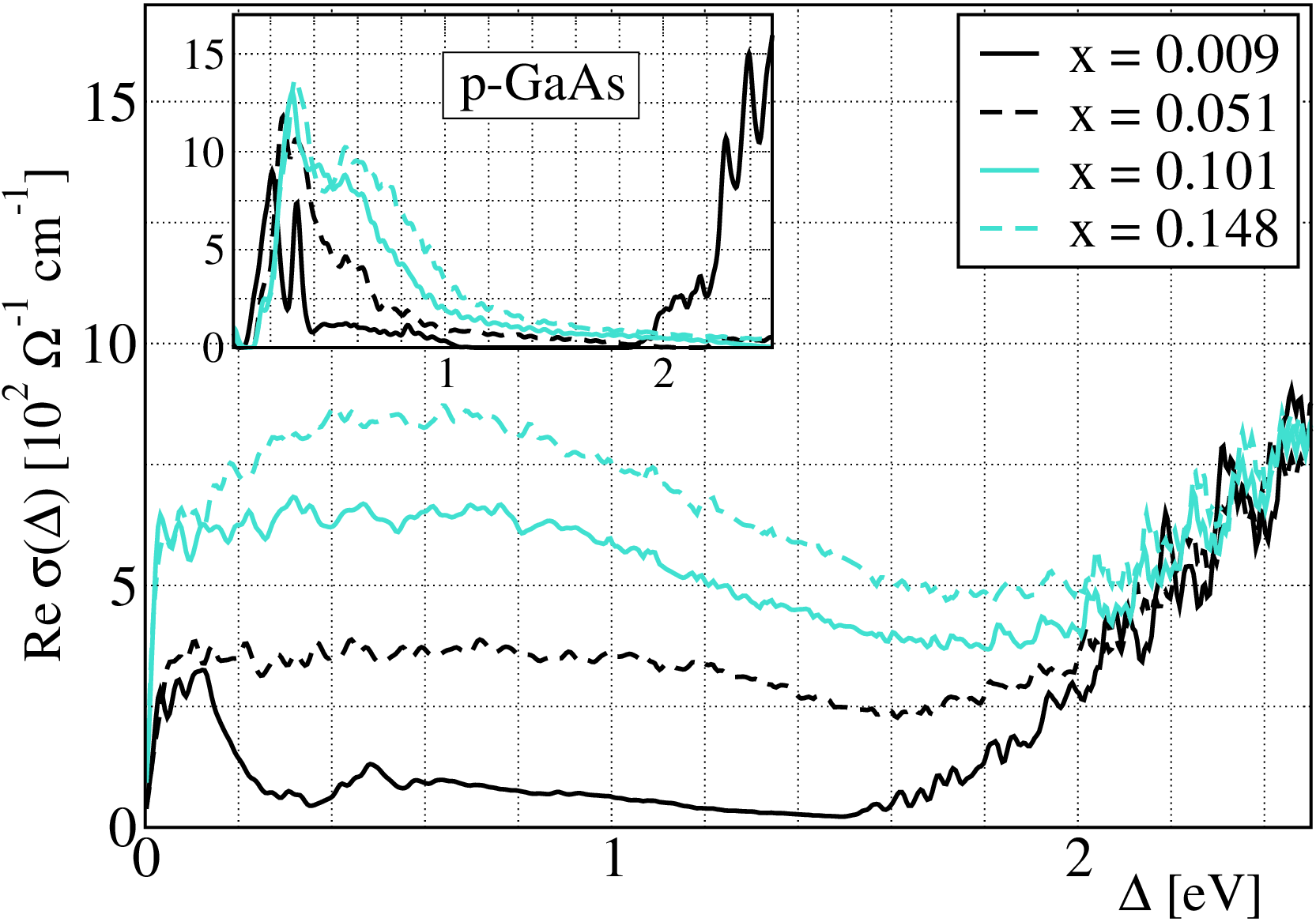}}
\caption{\label{fig:B_abs_conc}
 Model B -- Conductivity ${\rm Re} \, \sigma(\Delta,x)$ for
 various concentrations $x=0.009$ (686 atoms, 89 $\vec k$ vectors)
 and $x=0.051$, $0.101$, and $0.148$ (432 atoms, 240 $\vec k$ vectors).
 The results are averaged over 8 different disorder configurations.
 The inset shows the results for p-doped GaAs
 in the clean limit of the model with the corresponding number of holes
 added.
}
\end{figure}

The Figures~\ref{fig:A_abs_conc} and \ref{fig:B_abs_conc} show the
conductivity obtained from models A (Ma{\v s}ek) and B (Tang, Flatt{\'e}),
respectively. In either
case, the low frequency values are of limited accuracy as the system is of a
finite size which implies a finite minimal level spacing. The minimal
cut-off energy difference is set here as $\Delta_{\rm min} \sim 5$~meV
by using a broadened $\delta$-function of this width in Eq.~\ref{eq:conductivity}.

As there is a clear qualitative difference
between the two models let us start our
discussion with model A (Ma{\v s}ek). For this model we 
find a low-frequency Drude peak which has also been observed in
experiments.\cite{Burch06a} For low concentrations, $x \lesssim 0.01$,
additional peaks at $\approx 0.15$~eV and $\approx 0.33$~eV can
be identified clearly. These additional peaks correspond to inter-valence band
transitions which blue-shift to higher energy differences and become wider
as the number of impurities increases. This is very similar to
the experimental observations in p-doped GaAs.\cite{Songprakob02}
Eventually, for concentrations $x \gtrsim 0.05$, these peaks are
vanishing in the background of the Drude peak centered at $\Delta=0$.
The width of this Drude peak can be related to a scattering time $\tau$.
This width and the peak height can be estimated by fitting
a Lorentzian to the conductivity curve for small $\Delta$,\cite{Burch06a}
see Fig.~\ref{fig:A_abs_conc}. As this fitting procedure
can only be performed with some uncertainty the resulting
scattering times should be considered as qualitative estimates.
For the concentrations studied they
range from $\tau \approx 27$~fs to $6$~fs,
see Table~\ref{tab:scattering}. The saturation of the scattering time
at high concentration is due to the break down of the
independent scattering mechanism.
The height of the peak gives the DC-conductivity, e.g.,
$\sigma_{\rm DC}\sim 4000\; {\rm \Omega}^{-1} {\rm cm}^{-1}$ for $x=0.05$,
which is one order of magnitude larger than the experimentally
observed values.\cite{Katsumoto01,Hirakawa02,Burch06a}
Additionally the peak height and width are
related to an effective mass $m^*$. The estimation of this
effective mass is based on a sum rule for
the evaluation of the plasma frequency,
\begin{equation}
 \label{eq:plasma_frequency}
 \omega_p^2 = \frac{2}{\pi \hbar} \int\limits_0^{\Delta_{\rm max}}
 {\rm d}\Delta \, {\rm Re} \sigma_z(\Delta) =
 \frac{n_0 e^2}{m^*} \, ,
\end{equation}
with $n_0$ being the carrier density and $\Delta_{\rm max}$
the cut-off energy. When Eq.~\ref{eq:plasma_frequency} is
applied to the Lorentzian fit of the Drude peak the cut-off energy can
be set to infinity. This procedure of fitting a Lorentzian
and applying the sum rule leads to
effective masses $m^*/m_e \approx 0.9$, see Table~\ref{tab:masses}.
Based on the estimates for the
scattering times and these effective masses
we conclude that $(k_{\rm F} \, l)^* \approx 10$~--~$14$,
see Table~\ref{tab:scattering},
for $x=0.01$~--~$0.15$ which means that the assumption
of weak scattering is applicable. However, the
compensation of some of the Mn acceptors due to
intersticials lowers the effective hole density and
thus also decreases the values for $(k_{\rm F} l)^*$ accordingly.
The estimates based on experimental measurements lie
between $k_{\rm F} \, l \approx 0.75$ \cite{Hirakawa02,Wagner06}
and  $k_{\rm F} \, l \approx 1$~--~$5$\cite{Burch06a,Jungwirth07a}.
Finally, we give an estimate of
the mobility $\mu \approx 13\; {\rm cm}^2/{\rm Vs}$ for $x=0.1$
which is by a factor $3-5$ larger than the experimental
values.\cite{Hirakawa02,Burch06a}

A different approach based on the sum rule is to
apply Eq.~\ref{eq:plasma_frequency} directly to the 
numerical data and choose a cut-off that includes all 
inter-band transitions but excludes
valence band to conduction band transitions\cite{Sinova02},
i.e. $\Delta_{\rm max} = 1.5$~eV.
This means that not only
the narrow peak at $\Delta=0$ is considered but the wider overall
shape of $\sigma(\Delta)$ which decays on an energy scale $\sim 0.15$~eV.
This defines an optical mass $m_{\rm opt}$ which,
for model A (Ma{\v s}ek), remains approximately constant,
i.e. $m_{\rm opt}/m_e \approx 0.5$, for $x=0.01$~--~$0.15$,
see Table~\ref{tab:masses}.
This is to be compared with a stronger increase of the magnitude of the
optical masses $m_{\rm opt}/m_e=0.6$~--~$1.2$ which we obtain
by the same method for p-doped GaAs with the corresponding
number of holes added. This discrepancy not only in the magnitude
but also in the rate of the increase of $m_{\rm opt}/m_e$
with respect to the concentration $x$
indicates that the band structure of the host valence band is 
altered due to the disorder. 

In strong contrast to model A (Ma{\v s}ek), there is no low-frequency Drude peak
for model B (Tang, Flatt{\'e}). For low concentrations $x=0.009$ the impurity band
causes an absorption maximum around $0.1$~eV. Its half-width
of $\sim 50$~meV reflects the
width of the impurity band. For higher concentrations the impurity
band completely merges with the host valence band leading to a
more or less constant optical absorption. This
behavior is in qualitative agreement with the experimental observations in
Refs.~\onlinecite{Katsumoto01} and \onlinecite{Hirakawa02}.
At concentrations $x \sim 0.1$ a broad
feature with a maximum at $0.5$~eV forms due to the larger
number of impurity states which
are shifted into the host gap. The absolute values of the
conductivity lie between $100\; {\rm \Omega}^{-1} {\rm cm}^{-1}$
and $900\; {\rm \Omega}^{-1} {\rm cm}^{-1}$ for the various concentrations
which is in good agreement with the measurements.\cite{Katsumoto01,Hirakawa02,Burch06a}
Applying the sum rule (\ref{eq:plasma_frequency})
with $\Delta_{\rm max} = 1.5$~eV
which defines the optical mass we find $m_{\rm opt}/m_e \approx 0.4$ for
$x=0.009$ increasing to  $m_{\rm opt}/m_e \approx 0.9$ for $x=0.148$,
see Table~\ref{tab:masses}.
The corresponding values of p-doped GaAs range from
$m_{\rm opt}/m_e \approx 0.6$ to $m_{\rm opt}/m_e \approx 1.2$.
As we have described in the previous section, model B (Tang, Flatt{\'e})
is close to a metal-insulator transition
for the impurity concentrations considered.
This is reflected by the very broad peak implying very short scattering times
which are inconsistent with the weak scattering limit.
For the larger concentrations $x \gtrsim 0.1$ we estimate
the half-width of the broad maximum of ${\rm Re}\,\sigma(\Delta )$ to
be $\sim 0.5$~eV implying $\tau \sim 1$~fs. From this
follows that $\mu \sim 3\; {\rm cm}^2/{\rm Vs}$ and
$k_{\rm F} \, l \sim 2.5$ so that the mean free path
is on the order of the wavelength of the carriers.
Hence the Ioffe-Regel-limit is
reached and a Drude-Boltzmann description of the
conductivity cannot be expected to be a valid approximation any more.

\begin{table}
 \begin{tabular}{| l | c | c | c | c |} \hline
  Mn concentration x & 0.01 & 0.05 & 0.1 & 0.15 \\ \hline
  model A: $\tau [fs]$ & 27 & 7 & 6 & 6 \\
  model A: $(k_{\rm F} l )^*$ & 10 & 14 & 14 & 14 \\
  model A: $(k_{\rm F} l )_{\rm opt}$ & 6 & 17 & 18 & 14 \\
  model B: $(k_{\rm F} l )_{\rm opt}$ & -- & -- & 2.5 & 2.8 \\ \hline
 \end{tabular}
 \caption{\label{tab:scattering} Estimates for the dimensionless
 scattering length $k_{\rm F} l$. The values for $(k_{\rm F} l )^*$
 are based on a Lorentzian fit while $(k_{\rm F} l )_{\rm opt}$ was
 obtained by direct integration of the numerical data. The $(k_{\rm F} l)$ values for
 model B indicate that for the given range of parameters
 the Drude-Boltzmann theory is not unambiguously applicable. We have nevertheless
 included the values for comparison with experimental data.
 }
\end{table}

\begin{table}
 \begin{tabular}{| l | c | c | c | c |} \hline
  Mn concentration x & 0.01 & 0.05 & 0.1 & 0.15 \\ \hline
  model A: $m^*/m_e$ & 1.0 & 0.5 & 0.9 & 1.0 \\
  model A: $m_{\rm opt}/m_e$ & 0.5 & 0.4 & 0.5 & 0.5 \\
  model B: $m_{\rm opt}/m_e$ & 0.4 & 0.6 & 0.7 & 0.9 \\ \hline
 \end{tabular}
 \caption{\label{tab:masses} Estimates for the effective masses
 based on the sum rule (\ref{eq:plasma_frequency}).
 The values for $m^*/m_e$ have been obtained by the Lorentzian fit 
 to the Drude peak obtained for model A (Ma{\v s}ek). The masses
 $m_{\rm opt}/m_e$ were calculated by integrating the numerical
 data directly.
 }
\end{table}

In Ref. \onlinecite{Burch06a} the reasoning
for the existence of an impurity band was based
on the analysis of effective masses. However, the same analysis performed
for our simulation data does not give a useful tool to distinguish
the case without an impurity band from the one where an impurity band can still be
identified. Although the density of states and the localization
properties are very different for the two models A and B we find rather
similar values for the effective and optical masses, see Table~\ref{tab:masses}.

%%%%%%%%%%%%%%%%%%%%%%%%%%%%%%%%%%%%%%%%%%%%
\section{Conclusions}
\label{sec:conclusions}
%%%%%%%%%%%%%%%%%%%%%%%%%%%%%%%%%%%%%%%%%%%%
We have presented tight-binding studies of the electronic
and optical properties of ${\rm Ga}_{1-x}{\rm Mn}_x {\rm As}$
based on two different parameterizations. The
first parameterization due to Ma{\v s}ek (model A),
is based on first-principles
calculations. In this model, the Mn impurity is described
by a change in the on-site and hopping terms as well
as the additional inclusion of the d-orbitals. This
approach leads to certain qualitative changes in the
electronic properties, i.e., the density of states. However,
it does not give rise to the formation of an impurity band.
Instead, the inclusion of the disorder has a very similar
effect as the inclusion of an equally large amount of holes
without additional disorder. For the optical properties that
can be deduced from Ma{\v s}ek's model (A), we find the formation of
a Drude peak as a clear signature of the disorder.
This is in agreement with recent experiments. The
estimations for the dimensionless scattering length $(k_{\rm F} \, l)$ and
the mobility $\mu$ give reasonable results while the 
absolute value of the DC conductivity is significantly larger
than those observed in experiments. The inter-band transitions, that are also
identifiable for low concentrations $x<0.1$, blue-shift with
increasing amount of disorder which is in contradiction to the
experiments performed on GaMnAs.

The second parameter set, by Tang and Flatt{\'e} (model B),
is based on a more phenomenological 
modelling of the Mn impurities. It describes the bound level of
a single Mn impurity rather well and leads to a formation of an
impurity band in the density of states of
${\rm Ga}_{1-x}{\rm Mn}_x {\rm As}$ for small $x\lesssim 0.01$.
This impurity band merges
with the host valence band at concentrations $x\sim 0.01$. The Fermi
energy lies within this impurity band and the states at the Fermi level
are extended for Mn concentrations between 5\% and 15\%.
However, the eigenstates become localized for higher energies
closer to the band edge. The optical conductivity is characterized by
the absence of a Drude peak at zero frequency and a rather
featureless shape. It shows a broad maximum for larger
Mn concentrations with a half-width
of approximately $0.5$~eV. From this width we deduce
short scattering times leading to the conclusion
that a Drude-Boltzmann theory of weak scattering is not
applicable.

As the experiments on the electronic properties of
GaMnAs are not entirely conclusive at the moment
it is difficult to judge which model fits best for a given physical
quantity. Either model describes some of the experimental findings correctly while
it fails for others. 

%%%%%%%%%%%%%%%%%%%%%%%%%%%%%%%%%%%%%%%%%%%%
\section*{Acknowledgements}
\label{sec:acknowledgements}
%%%%%%%%%%%%%%%%%%%%%%%%%%%%%%%%%%%%%%%%%%%%
The authors thank J.~Ma{\v{s}}ek
for helpful discussions on his model, T.~Dietl for useful
comments before submission of the manuscript and B.~L.~Gallagher
for providing some new experimental data. This work was supported by
the SFB 689.

%%%%%%%%%%%%%%%%%%%%%%%%%%%%%%%%%%%%%%%%%%%%

%\bibliography{GaMnAs}

%\begin{references}
%%\begin{thebibliography}
%%\bibitem[*]{MT_email}
%% email-address: {\tt marko.turek@physik.uni-regensburg.de}
%\bibitem{Burch06a} K.S.~Burch and D.B.~Shrekenhamer and E.J.~Singley, cond-mat0603851
% (2006). 
%%\end{thebibliography}
%\end{references}

\end{document}